%% file: LeakyPick.tex
\documentclass[sigconf, nonacm, hidelinks]{acmart}
\usepackage{tikz}
\usepackage{amsmath}
  
\usepackage{xspace}
\usepackage{hyperref}

\usepackage{xcolor}
\usepackage{pifont}
\usepackage[utf8]{inputenc}
\usepackage{multirow}

\usepackage{fancyhdr}
\usepackage{url}

\newcommand{\changed}[1]{#1}
\newcommand{\morechange}[1]{#1}

\usepackage{amssymb}

\input{macros}


\begin{document}
%-------------------------------------------------------------------------------

\fancyhead[HLE, HRO]{\bfseries Pre-print of paper to be published at ACSAC2020\\acmDOI: 10.1145/3427228.3427277}

%don't want date printed
\date{}

% make title bold and 14 pt font (Latex default is non-bold, 16 pt)
\title{\ourname: IoT Audio Spy Detector}

%for single author (just remove % characters)
%\author{
%{\rm }\\
%Technische Universität Darmstadt
%\and
%{\rm Anna Pazii}\\
%Second Institution
% copy the following lines to add more authors
%\and
%{\rm Markus Miettinen}\\
%Technische Universität Darmstadt
%\and
%{\rm William Enck}\\
%North Carolina State University
%\and
%{\rm Ahmad-Reza Sadeghi}\\
%Technische Universität Darmstadt
%} % end author

\author{Richard Mitev}
\email{richard.mitev@trust.tu-darmstadt.de}
\affiliation{%
  \institution{Technical University of Darmstadt}
}

\author{Anna Pazii}
\email{anna.pazii@inria.fr}
\affiliation{%
  \institution{University of Paris Saclay}
}

\author{Markus Miettinen}
\email{markus.miettinen@trust.tu-darmstadt.de}
\affiliation{%
  \institution{Technical University of Darmstadt}
}
\author{William Enck}
\email{whenck@ncsu.edu}
\affiliation{%
  \institution{North Carolina State University}
}
\author{Ahmad-Reza Sadeghi}
\email{ahmad.sadeghi@trust.tu-darmstadt.de}
\affiliation{%
  \institution{Technical University of Darmstadt}
}

%-------------------------------------------------------------------------------
\begin{abstract}
\input{abstract}
%-------------------------------------------------------------------------------
\end{abstract}

%\keywords{privacy, IoT, voice assistants, machine learning}

\maketitle

%\IEEEpeerreviewmaketitle

%-------------------------------------------------------------------------------
\section{Introduction}
\label{sect:intro}
\input{intro}
%-------------------------------------------------------------------------------

%-------------------------------------------------------------------------------
%\section{Motivation}
%\label{sect:motivation}
%\input{motivation}
%-------------------------------------------------------------------------------

%-------------------------------------------------------------------------------
\section{Background}
\label{sect:preliminaries}
\label{sect:background}
\input{preliminaries}
%-------------------------------------------------------------------------------

%-------------------------------------------------------------------------------
\section{Solution Overview}
\label{sect:overview}
\input{overview}
%-------------------------------------------------------------------------------

%-------------------------------------------------------------------------------
\section{\ourname Design}
\label{sect:system}
\input{system}
%-------------------------------------------------------------------------------

%-------------------------------------------------------------------------------
\section{Evaluation}
\label{sect:evaluation}
\input{evaluation}
%-------------------------------------------------------------------------------

%-------------------------------------------------------------------------------
%\section{Hidden Wake Words}
%\label{sect:hiddenwakewords}
%\input{hiddenwakewords}
%-------------------------------------------------------------------------------

%-------------------------------------------------------------------------------
\section{Discussion}
\label{sect:discussion}
\input{discussion}
%-------------------------------------------------------------------------------

%-------------------------------------------------------------------------------
\section{Related Work}
\label{sect:relatedwork}
\input{relatedwork}
%-------------------------------------------------------------------------------

%-------------------------------------------------------------------------------
\section{Conclusion}
\label{sect:conclusion}
\input{conclusion}
%-------------------------------------------------------------------------------

%-------------------------------------------------------------------------------
\bibliographystyle{ACM-Reference-Format}
\bibliography{LeakyPick}

%-------------------------------------------------------------------------------
% Moved to body of text
%\appendix
%\section*{Appendix}
%\label{sect:appendix}
%\input{appendix}
%-------------------------------------------------------------------------------

%%%%%%%%%%%%%%%%%%%%%%%%%%%%%%%%%%%%%%%%%%%%%%%%%%%%%%%%%%%%%%%%%%%%%%%%%%%%%%%%
\end{document}

%% file: macros.tex
\usepackage{comment}
\excludecomment{longversion}
\newcommand{\ourname}{LeakyPick\xspace}

\usepackage{makecell}
\newcommand{\thickhline}{\Xhline{2\arrayrulewidth}}

\newcommand{\myparagraph}[1]{\vspace{0.25em}\noindent\textbf{#1:}}
\newcommand{\myparagraphi}[1]{\vspace{0.25em}\noindent\textit{#1:}}

\definecolor{Medium Carmine}{rgb}{0.671, 0.251, 0.188}
\definecolor{Clover}{rgb}{0.213, 0.304, 0.109}
\definecolor{Hot Toddy}{rgb}{0.663, 0.476, 0.204}

%% file: abstract.tex
% Area
Manufacturers of smart home Internet of Things (IoT) devices are increasingly adding voice assistant and audio monitoring features to a wide range of devices including smart speakers, televisions, thermostats, security systems, and doorbells.
% Problem
Consequently, many of these devices are equipped with microphones, raising significant privacy concerns: users may not always be aware of when audio recordings are sent to the cloud, or who may gain access to the recordings.
% Solution
In this paper, we present the \ourname architecture that enables the detection of the smart home devices that stream recorded audio to the Internet \morechange{in response to observing a sound}.
%without the user's consent.
Our proof-of-concept is a \ourname device that is placed in a user's smart home and periodically ``probes'' other devices in its environment and monitors the subsequent network traffic for statistical patterns that indicate audio transmission. 
Our prototype is built on a Raspberry Pi for less than USD \$40 and has a measurement accuracy of 94\% in detecting audio transmissions for a collection of 8 devices with voice assistant capabilities.
Furthermore, we used \ourname to identify 89 words that an Amazon Echo Dot misinterprets as its wake-word, resulting in unexpected audio transmission.
%\richard{With this approach, we found 89 words Echo Dot devices misinterpret for their wake-word.} % Methodology
% Results
%We further extended the \ourname architecture to ``fuzz'' the wake-words of the voice assistants and discovered 135 hidden wake-words that unexpectedly cause audio recordings to be streamed to the cloud, while bearing little to no phonetic similarity to the ``official'' wake-word.
% Take-away
\ourname provides a cost effective approach \morechange{to help} regular consumers monitor their homes for \morechange{sound-triggered devices that unexpectedly transmit audio to the cloud.}
%unexpected audio transmissions to the cloud.

%% file: intro.tex
%!TeX root=LeakyPick.tex

% Many IoT devices are equipped with audio interface (typically a microphone)
% some with obvious purpose such as voice assistant and some less obvious such
% as a surveillance camera (e.g., from Ring).
Consumer Internet of Things (IoT) devices have emerged as a promising technology to enhance home automation and physical safety.
While the smart home ecosystem has a sizeable collection of automation platforms, market trends in the US~\cite{Tofel2018} suggest that many consumers are gravitating towards Amazon Alexa and Google Home.
These two platforms are unique from the other automation platforms (e.g., Samsung SmartThings, WeMo) in that they focused much of their initial smart home efforts into smart speaker technology, which allows users to speak commands to control smart home devices (e.g., light switches), play music, or make simple information queries.
This dominance of Amazon Alexa and Google Home might suggest that consumers find voice commands more useful than complex automation configurations.

% Microphones in the home is a privacy problem
For many privacy-conscious consumers, having Internet connected microphones scattered around their homes is a concerning prospect.
This danger was recently confirmed when popular news media reported that Amazon~\cite{BusinessInsider2019}, Google~\cite{VRTNews2019,Google2019}, Apple~\cite{TheGuardian2019}, Microsoft~\cite{Motherboard2019}, and Facebook~\cite{Bloomberg2019-2} are all using contractors to manually analyze the accuracy of voice transcription.
The news reports include anecdotes from contractors indicating they listened to many drug deals, domestic violence, and private conversations.
Perhaps more concerning is that many of the recordings were the result of false positives when determining the ``wake-word'' for the platform.
That is, the user never intended for the audio to be sent to the cloud.

% For some devices the audio interface is “hidden” to the user.
Unfortunately, avoiding microphones is not as simple as not purchasing smart speakers.
Microphones have become a pervasive sensor for smart home devices.
For example, it is difficult to find a smart television that does not support voice controls via the display or the handheld remote control.
Smart thermostats (e.g., Ecobee) commonly advertise that they have dual function as a smart speaker.
Surveillance cameras (e.g., Ring, Wyze) are designed to notify users of events, but are in fact always recording.
Perhaps most concerning was the report that the Nest security system includes a microphone~\cite{BusinessInsider2019}, despite no packing material or product documentation reporting its existence.
While the manufacturers of these devices might argue that users can disable microphone functionality in device software,
history has repeatedly demonstrated that software can and will be compromised.
Furthermore, mass collection and storage of audio recordings increases concerns over the potential for a ``surveillance state''
(e.g., Ring has recently been criticized for working with local police~\cite{Haskins2019}).
%Furthermore, changes in company policies may result in ``legitimate'' recording of audio or storage of audio.

% Our research question: How can a user effectively detect that a device is
% using the microphone to capture the user’s voice and record them without the
% user’s knowledge. In this context the device can either be compromised by an
% adversary (including nation state) or the recording is a benign but hidden
% feature and principally privacy violating. This also includes cases where
% specific words trigger the devices (and the recording) when these words are
% of special interest for national security.
%Our research seeks to answer the question: \textit{Can a user effectively detect if a device is using the microphone to capture the user's voice and record them without their knowledge?}
Our research seeks to answer the question: \morechange{\textit{Can a user effectively detect if a smart home device expectantly transmits audio recordings to Internet servers in response to a sound trigger?}}
\changed{Such failures can occur in two types of situations:
(1)~devices that are not expected to have recording capability or are expected to have the capability disabled transmit user audio, or,
(2)~devices that are expected to have recording capability enabled, but transmit audio in response to unexpected stimuli (e.g., unexpected or misunderstood wake-words).}
%\changed{We answer this question by identifying devices which have a microphone and react on audio cues without the user noticing he has a microphone-enabled device and to what words it reacts to, including wake-words it should not reat to, and alerting the user about this.}
\changed{In both cases} we are primarily concerned with the benign, but hidden, recording and immediate transmission of audio to cloud services, as such behaviors can potentially lead to mass surveillance. 
%\changed{This means, in particular, devices recording and streaming audio without the proper wake-word being used to activate the voice assistant. In most cases this will be due to misactivations of the device.}
%We also seek to identify the existence of compromised devices that immediately transmit audio to the network as it is recorded.
%\todocite{better remove this following sentence about things we cannto do}
%However, we recognize that more sophisticated attacks could buffer audio and use covert methods to exfiltrate recordings in ways that circumvent our detection.
We believe there is significant utility in detecting this subset of misbehaving devices,
%that immediately transmit audio to the network, 
particularly given the limited storage capacity of many low-level IoT devices.
\morechange{Additionally, we only consider audio transmission that occur in response to a sound-based trigger event.
Devices that continuously stream audio are detectable by monitoring bandwidth use.}
%Random intermittent recordings are less likely to contain sensitive information.}

% seems awkward to forward reference related work in the intro. we should just state the primarily work here
%As we will elaborate on related work in Section~\ref{sect:relatedwork}, 
Existing approaches for identifying unintended data transmissions out of the user's network~\cite{liu2018detecting,cheng2018dewicam} focus on other modalities (e.g., video) and rely on assumptions that do not apply to audio transmissions (e.g., some devices require an utterance of specific wake-words). Furthermore, while traffic analysis approaches targeting IoT devices have been proposed~\cite{DBLP:journals/corr/abs-1804-07474, Sivanathan2017CharacterizingAC}, to the best of our knowledge there are no earlier approaches specifically targeting traffic of microphone-enabled IoT devices. Additionally, prior work attacking voice assistants and voice recognition~\cite{diao2014your, alepis2017monkey, vaidya2015cocaine, carlini2016hidden, carlini2018audio, kumar2018skill, zhang2017dolphinattack, zhang2018understanding, schonherr2018adversarial, yuan2018commandersong, mitev2019alexa} focuses on maliciously issuing commands or changing the interaction flow without the victim noticing.
% In this paper ...

In this paper, we present the \ourname architecture, which includes a small device that can be placed in various rooms of a home to detect the existence of smart home devices that stream recorded audio to the Internet.
%\changed{Our key insight is that unexpected recording behavior is generally not contextual and can be triggered at any time.}
\ourname operates by periodically ``probing'' an environment (i.e., creating noise) 
%when the user is not home (to avoid annoyance) 
and monitoring subsequent network traffic for statistical patterns that indicate the transmission of audio content. %To detect a device sending audio to the cloud while the user is at home no audio probing is necessary.}
\changed{
By using a statistical approach, \ourname's detection algorithm is
generalizable to a broad selection of voice-enabled IoT devices, eliminating the need for time-consuming training required by machine learning.
}

\changed{We envision \ourname being used in two scenarios.
First, \ourname can identify devices for which the user does not know there is a microphone, as well as smart home devices with smart speaker capabilities (e.g., an Ecobee thermostat) that the user was not aware of, or thought were disabled (e.g., re-enabled during a software update).
%Note that \ourname needs to include common wake-words when probing to handle the latter case.
%We evaluated this scenario by simulating an environment when the user expects that the smart speaker capabilities of a device are disabled, but they are not.
To evaluate this scenario, we studied eight different microphone-enabled IoT devices and observed that
that \ourname can detect their audio transmission with 94\% accuracy.
}
\changed{
Second, \ourname can be used to determine if a smart speaker transmits audio in response to an unexpected wake-word.
To evaluate this scenario,}
we used \ourname to perform a wake-word fuzz test of an Amazon Echo Dot, discovering 89 words that unexpectantly stream audio recordings to Amazon.
\changed{
For both scenarios, \ourname can run when the user is not home (to avoid annoyance), since this behavior is generally not contextual the users' presence.
}
%, as discussed in Sect.~\ref{sect:traffic-profiling}.
%making resource and time-consuming generation of device-type-specific detection profiles as outlined in Sect.~\ref{sect:evaluation} unnecessary.
%This suggests that
%In doing so, \ourname provides a cost effective approach for regular consumers to monitor their homes for unexpected audio transmissions to the cloud.

% As a side-effect of this investigation we found out that voice assistants, in
% our case the highly popular Amazon echo, can be triggered by wake words not
% included in the public list of wake words provided by the vendor -- a slight
% evidence that this feature can be misused because it is not under the user’s
% control.
%As a side-effect of our investigation of voice assistants, we discovered several non-``wake-words'' that consistently cause the voice assistants to send audio recordings to the cloud.
%This discovery led us to ``fuzz'' (cf. Sect.~\ref{sect:overview-fuzzing}) the voice assistants to systematically identify more hidden wake-words.
%In doing so, we identified 135 words that trigger the voice recognition functionality on Alexa, the most popular voice assistant at the time of writing. Somewhat surprisingly, we concluded that almost all words that most reliably trigger a reaction, have a phonetic distance of \emph{50\% or more} to the actual wake-word (cf. Sect.~\ref{sect:fuzzing}).

This paper makes the following contributions: 

\begin{itemize}
  \item \textit{We present the \emph{\ourname} device for identifying smart home devices that unexpectedly record and send audio to the Internet \morechange{in response to observing a sound}.}
    The device costs less than USD \$40 and can be easily deployed in multiple rooms of a home.

  \item \changed{\textit{We present a novel audio-based probing approach for estimating the likelihood that particular devices react to audio.}}
    Our approach has a 94\% accuracy for a set of devices known to transmit audio to the cloud.
    We also show that the approach is generalizable to different device types without the need to pre-train costly device-type-specific detection profiles.
    
  %\item Based on an analysis with \textit{several real-world voice-controlled IoT devices in real deployment settings}, we show that the approach is \textit{generalizable to different device types} without the need to pre-train costly device-type-specific detection profiles.
  
  \item \textit{We show that \emph{\ourname} can identify hidden wake-words that cause unexpected audio transmission.}
    Our analysis of an Amazon Echo Dot identified 89 incorrect wake-words.
  %\item \textit{We discover 135 ``hidden'' wake words in popular voice assistants.}
  %  To do so, we performed a systematic investigation that ``fuzzes'' a device to determine which words consistently result in sending audio recordings to the cloud.

\end{itemize}

%There exist approaches for injecting probes in an inaudible manner utilizing ultrasound audio~\cite{zhang2017dolphinattack, song2017inaudible}, but as these are dependent on the specific technical characteristics of the targeted devices (that are by default not known beforehand), we limit \ourname to the use of human-audible signals.

Finally, \ourname uses human-audible noises, which may be annoying to physically present users.
Prior work has suggested the use of inaudible sound to control voice assistants using ultrasound audio~\cite{zhang2017dolphinattack, song2017inaudible}.
However, these approaches are specific to the technical characteristics of the targeted devices.
Therefore, they are not immediately applicable to our goal of identifying unknown devices streaming audio.
We leave the task of creating generic models of transmitting audio via ultrasonic sound as a topic for future work.

The remainder of this paper proceeds as follows.
Section~\ref{sect:preliminaries} provides background on devices with voice control and audio interfaces. 
Section~\ref{sect:overview} overviews our architecture. 
Section~\ref{sect:system} describes our design and implementation.
Section~\ref{sect:evaluation} evaluates the accuracy of \ourname.
%Section.~\ref{sect:hiddenwakewords} presents our analysis of hidden wake words.
Section~\ref{sect:discussion} discusses our approach and security considerations. 
Section~\ref{sect:relatedwork} overviews related work.
Section~\ref{sect:conclusion} concludes.

%\todocite{write about we tried ML and we tried inaudible injecting with ultrasonic}

%% file: preliminaries.tex
%!TeX root=LeakyPick.tex

IoT devices increasingly use audio sensing for enabling voice-based control by the user or for other audio-based use cases.
Examples of such devices include smart audio security systems~\cite{hivehub360},
smart audio event-detecting IP cameras~\cite{eventdetecting},
vacuum cleaner robots equipped with microphones and nightvision~\cite{vacuumcleaner}, and smart fire alarms with a self-testing siren~\cite{siren}.
Due to the abundance of devices with audio sensing capabilities, the user may not always be aware of when a particular device will record audio and send it to the cloud.
Sending audio to the cloud is frequently required for voice-control based user interfaces, as speech-to-text translation often needs more computational resources than are available on IoT devices.
%typically performed in the back-end system of the service provider in order to apply more advanced speech processing algorithms that can not be executed on simple IoT devices. 

\begin{figure}[t]
	\centering
	\includegraphics[trim=10.7cm 10.5cm 15cm 2.75cm, clip, width=.35\textwidth]{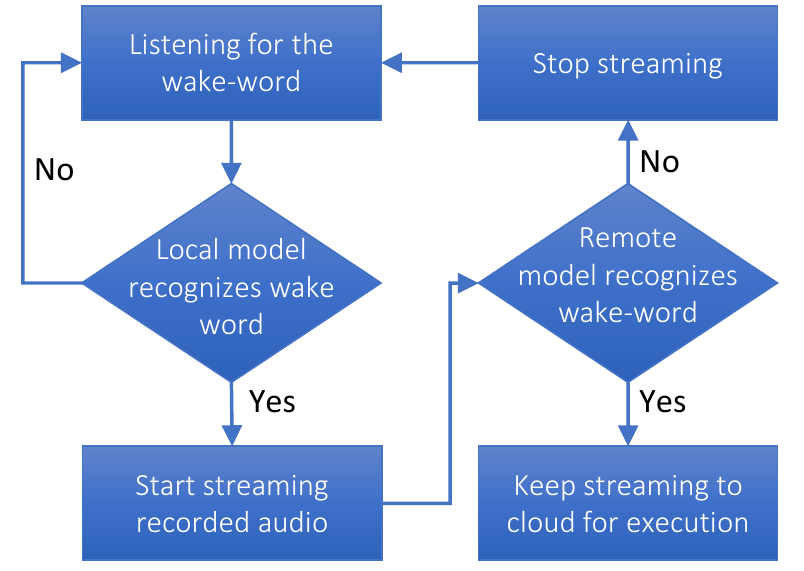}
	\caption{Overview of wake-word detection process in voice-controlled devices such as the Amazon Echo}
	\label{fig:alexa_recognition}
\end{figure}

Devices with voice-control interfaces typically use local speech recognition for detecting a specific set of ``wake-words'' (i.e., utterances meant to be used by the user to invoke the voice-control functionality of the device).
When the local model detects the utterance of a potential wake-word, the device starts sending audio to back-end servers for voice-to-text translation.
In order to not miss speech commands uttered by users, the local model needs to be configured to recognize any utterance that resembles the intended wake-word.
In case of the Alexa voice assistant, it is then the task of the back-end service to verify whether the observed utterance really represents a wake-word or not, as it is equipped with a more comprehensive speech recognition model and is not limited by the potentially scarce computational capabilities of the IoT device recording the audio.
%\will{We need to tweak this paragraph. it starts out by saying that there is lots of recording (good for us) and then says, oh, vendors really can't record all the time. This is accurate, but not the motivation we want for the new version of the paper that is focused on monitoring}
%
Figure~\ref{fig:alexa_recognition} overviews this typical approach. 
% Will: paragraph seemed redundant with the last
%The device continuously records audio and feeds this to the local speech detection model. 
%If the local model triggers a detection of a wake-word, the device will stream the recorded audio to the back-end system for further speech-to-text processing. 
%The back-end system will then utilize its own wake-word detection model to analyze the streamed audio input and verify whether the wake-word detection of the local model was correct or not. 
%If the streamed audio is determined to contain the wake-word, it is sent to the speech-to-text system for further processing. Otherwise, the back-end system will cancel command processing and instruct the device to stop recording user audio. 
%%As we will discuss in more detail in Sect.~\ref{sect:evaluation}, we utilize this wake-word detection logic to distinguish how likely specific utterances are mistaken as the wake-word by different components of the audio detection logic.

Problems arise when the local or online detection model mistakenly classifies specific audio inputs as the wake-word and consequently starts sending the recorded audio to the cloud, thereby potentially leaking sensitive information. 
% Will: next two sentences are redundant with the previous paragraph
%If the offline model recognizes a wake-word, this small audio snippet will be sent to the cloud for further processing. If the cloud accepts this wake-word, consecutive audio will be streamed to the cloud for command detection. 
Private information may also be leaked unintentionally when the user is unaware that a device will react to specific wake-words.
%and thereby may inadvertently trigger the audio-recording functionality without noticing it.

%Many smart devices determine based on recorded audio whether they should or should not execute an operation. For the user it may not be clear what kind of audio signals trigger particular operations as the machine learning models used in audio detection are typically proprietary, i.e., not public, making public scrutiny of their behavior difficult if not impossible. In some cases the manufacturer may even not be willing to disclose the intended functionality of devices, e.g., in the case of audio-based burglary alarms, in order not to disclose sensitive information making it easier for malicious persons to bypass alarms. Because of this a device may react to audio events the the user would not expect it to. The inherent unavailability of transparent information makes it therefore difficult for users to fully understand to what types of audio inputs a particular device may react to. As many device use a proprietary algorithm or machine learning model in the cloud, this audio sample will be sent to a server and the user looses control of their private data.

Finally, attacks targeting voice assistants can use malicious audio signal injection to trick the assistant to perform actions. 
In these attacks, the adversary either waits for the user to be asleep or not present~\cite{diao2014your,alepis2017monkey} or uses inaudible~\cite{zhang2017dolphinattack,song2017inaudible,mitev2019alexa} or unrecognizable~\cite{schonherr2018adversarial,carlini2016hidden,vaidya2015cocaine,yuan2018commandersong} signals to stage the attack, making it very difficult for the victim user to realize that the device is being attacked.

Our goal is to provide tools that enable users to \changed{detect and identify 1) devices that are not expected to have audio recording transmission capability, 2) devices for which audio recording transmission is expected to be disabled, and 3) unexpected wake-words that cause devices to unexpectedly transmit audio recordings.}

\myparagraph{Threat Model and Assumptions}
%\todocite{extend, write that buffering attack dont scale}
In this paper, we are concerned with threats related to IoT devices that stream recorded audio over the Internet using Wi-Fi or a wired LAN connection in response to audio signals recognised by the device as potential voice commands or different sounds the device reacts to. 
As processing voice commands is (except for the detection of the device's wake-word) implemented on the server-side, we assume that recorded audio is transmitted instantaneously to the cloud to allow the voice-controlled device to promptly react to user commands. 
%Fig.~\ref{fig:alexa_recognition}.
%Delaying the transmission of data or performing batch uploads is not possible for many devices, as they do not have enough memory, storage and computational power to either store a large amount of audio or analyze it locally.
%Due to this we assume devices to send the recorded audio to respective back-end servers for speech recognition as quickly as possible to minimize the delay between possible user command invocation and response.
We consider three main threat scenarios:
%related to such voice-controlled devices: 
\begin{enumerate}
    \item \changed{Benign IoT devices that may have undocumented microphones and audio-sensing capabilities, devices for which audio-sensing capabilities are unexpectedly enabled (e.g., via a software update), or devices whose wake-word detection is inaccurate, leading to audio being sent to the cloud without the users intent and knowledge.}
    \item Application-level attacks that cause a benign IoT device to send audio without the user's knowledge.
    For example, the Amazon Echo contained a vulnerability~\cite{echo-snoop} that allowed a malicious skill to silently listen.
    More recently, security researchers have reported~\cite{alexa-eavesdrop} the existence of eavesdropping apps targeting Alexa and Google Assistant. \changed{Note that in this scenario, the IoT device is benign, but it supports third-party applications that may be malicious.}
    %\item Audio events or utterances inadvertently recognized by the device as the device's wake-word, resulting in the \emph{unwanted streaming of audio recordings} to the device's back-end system for further processing. The main threat in this scenario is the \emph{unintentional disclosure} of potentially sensitive audio to the back-end system of the device's vendor. 
    \item Active attacks by an external adversary where the adversary tricks the targeted device to initiate transmission of audio data by \emph{injection of audio signals} to the device's environment so that the device will identify these as its wake-word. The main threat in this scenario is the \emph{unauthorized invocation} of device or service functionalities.
\end{enumerate}

%Our threat model primarily considers benign IoT devices, i.e., devices not compromised by an adversary. \changed{This includes devices that may have undocumented microphones and audio-sensing capabilities, or, devices whose wake-word detection is  inaccurate, leading to audio being sent to the cloud without the users intent and knowledge. However, our threat model also covers attacks where the adversary injects audio signals into the context of the voice assistant in order to trigger specific functionalities.}
We do not consider scenarios in which an IoT device is modified by the adversary to transform the device to act as a bugging device that records and stores audio locally without sending it to the network. 
While feasible, such attacks are much less scalable, as they must be tailored for each targeted device and are only applicable for devices with sufficient storage space.
We note, however, that \ourname is applicable to settings where an adversary has compromised an IoT device and immediately transmits audio data.
%Our approach does not consider malicious bugging devices as these store audio recordings locally on a high-capacity memory card which the attacker has to manually retrieve to obtain access to the recordings. Such highly targeted attacks require considerable effort from the adversary as well as physical access to the home of the victim, and are therefore outside the scope of our system.

%% file: overview.tex
%!TeX root=LeakyPick.tex
%\todocite{if possible, add more technical details}
The goal of this paper is to devise a method for regular users to reliably identify IoT devices that 1) are equipped with a microphone, 2) send recorded audio from the user's home to external services without the user's awareness, \changed{and 3) do so unexpectantly \morechange{in response to observing a sound} (e.g., unexpected wake-word, software update re-enabling smart speaker functionality).}
%\changed{An example of a scenario where \ourname is beneficial, is a smart home with new devices (e.g., a smart TV). \ourname will then audio probe the devices in the smart home when the user is away, identify if devices react to these audio cues and alert the user to which words the device reacts to. When the user is at home \ourname monitors the traffic of the device passively in the background (without audio probing) to detect if the device sends audio to its cloud and notify and potentially warn the user about such occurrences.}
If \ourname can identify which network packets contain audio recordings, it can then inform the user which devices are sending audio to the cloud, as the source of network packets can be identified by hardware network addresses. 
%This provides a way to identify both unintentional transmissions of audio to the cloud, as well as above-mentioned attacks (discussed in detail in Section~\ref{sect:relatedwork}), where adversaries seek to invoke specific actions by injecting audio into the device's environment.
Achieving this goal requires overcoming the following research challenges:
%
%1) Cannot constantly monitor devices. (this is why we do random probing. Can we make claims about how we do the probing?)
%2) Device traffic is encryption. We can't just look for audio codecs
%3) There are many types of devices. We need to find generic approaches that will work with previously unseen devices.
%4) Wake words are not known. Therefore, we opportunistically fuzz with different dictionary words over time.
%5) devices may perform local processing
%
\begin{itemize}

    \item \textit{Device traffic is often encrypted.}
    A na\"{i}ve solution that simply looks for audio codecs in network traffic will fail to identify most audio recordings.
    
    \item \textit{Device types are not known a priori.} 
    Devices transmit audio in different ways.
    We need to identify generic approaches that work with previously unseen devices.
    
\end{itemize}

Due to these challenges, our solution cannot passively monitor network traffic with the goal of differentiating the transmission of audio recordings from other network traffic.
While prior approaches such as HomeSnitch~\cite{oconnor2019homesnitch} are able to classify the semantic behavior of IoT device transmissions (e.g., voice command), they require \emph{a priori} training for each manufacturer or device model.
Since we seek to identify this behavior for potentially unknown devices, we cannot rely on supervised or semi-supervised machine learning.
%Therefore, \ourname takes an active approach to detecting transmission of audio recordings to cloud services.

\begin{figure}[t]
	\centering
	\includegraphics[trim=5.4cm 7.1cm 6.2cm 2cm, clip, width=0.9\columnwidth]{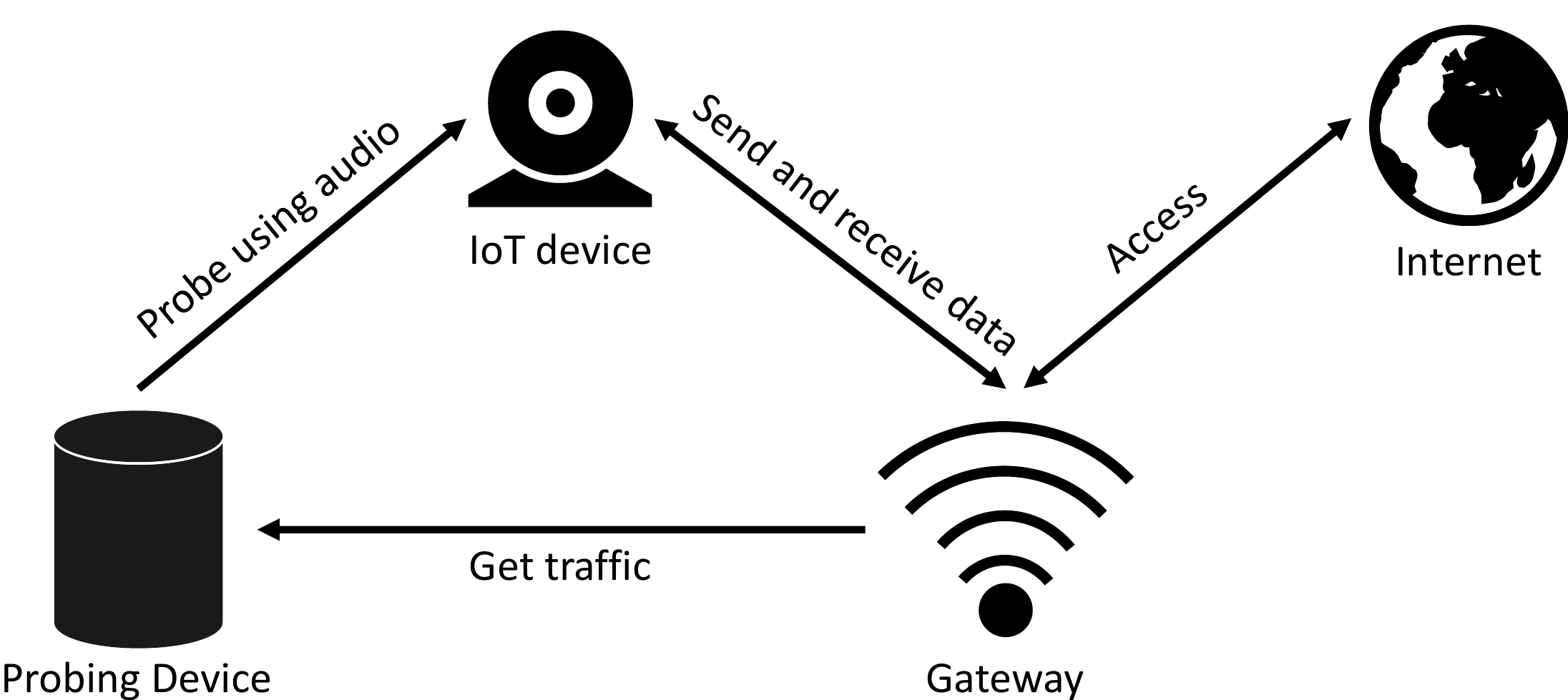}
	\caption{System set-up of \ourname}
	\label{fig:systemsetup1}
\end{figure}

At a high level, \ourname overcomes the research challenges by periodically transmitting audio \changed{(potentially prepended with wake-words)} into a room and monitoring the subsequent network traffic from devices.
As shown in Figure~\ref{fig:systemsetup1},
\ourname's main component is a probing device that emits audio probes into its vicinity.
%and subsequently monitor outgoing network traffic data sent by devices connected to the user's home network.
By temporally correlating these audio probes with observed characteristics of subsequent network traffic, \ourname identifies devices that have potentially reacted to the audio probes by sending audio recordings.
%out from the local network.
%Figure~\ref{fig:systemsetup1} shows the high level operation of \ourname.

%\ourname addresses these challenges by periodically probing an environment with human-audible noises.
%\ourname addresses the first challenge (encryption) by monitoring traffic characteristics.
\ourname identifies network flows containing audio recordings using two key ideas.
First, it looks for traffic bursts following an audio probe.
Our observation is that voice-activated devices typically do not send much data unless they are active.
For example, our analysis shows that when idle, Alexa-enabled devices periodically send small data bursts every 20 seconds, medium bursts every 300 seconds, and large bursts every 10 hours.
We further found that when it is activated by an audio stimulus, the resulting audio transmission burst has distinct characteristics.
However, using traffic bursts alone results in high false positive rates.

Second, \ourname uses statistical probing.
Conceptually, it first records a baseline measurement of idle traffic for each monitored device.
Then it uses an \emph{independent two-sample t-test} to compare the features of the device's network traffic while being idle and of traffic when the device communicates after the audio probe.
%In doing so, \ourname can be more certain that the device transmits data in response to audio.
This statistical approach has the benefit of being inherently device agnostic.
As we show in Section~\ref{sect:evaluation}, this statistical approach performs as well as machine learning approaches, but is not limited by \emph{a priori} knowledge of the device. It therefore outperforms machine learning approaches in cases where there is no pre-trained model for the specific device type available.

Finally, \ourname works for both devices that use a wake word and devices that do not.
For devices such as security cameras that do not use a wake word, \ourname does not need to perform any special operations.
Transmitting any audio will trigger the audio transmission. 
To handle devices that use a wake word or sound, e.g., voice assistants, security systems reacting on glass shattering or dog barking, \ourname is configured to prefix its probes with known wake words and noises (e.g., "Alexa", "Hey Google").
It can also be used to fuzz test wake-words to identify words that will unintentionally transmit audio recordings.
%\will{really? does it need to be a word in a specific language? Is there language detection? What about Echo devices that identify home intrusion by listening for the sound of broken class?}
%To handle devices that use a wake word, \ourname is configured to prefix its probes with range of known wake words (e.g., "Alexa", "Hey, Google", "Hey, Siri").
%\will{To what extent do we already do this?}
%We note that \ourname could also be used to fuzz devices to identify hidden wake words.
%However, we leave this investigation to future work.
%\will{Do we want to say this?}

%% file: system.tex
%!TeX root=LeakyPick.tex

This section describes the central aspects of \ourname's design.
We primarily focus on audio event detection. 
We then describe our system implementation on a Raspbery Pi 3B.

% Will: let's drop this subsection for the ACSAC submission, since we are too long
\begin{longversion}
\subsection{System Design}
\ourname is designed to work in a smart home living room environment with multiple audio-enabled IoT devices. The probing device is placed in the room where the user seeks to identify which IoT devices react to audio and stream recordings to the Internet. 
The probing device accomplishes this task by injecting audio probes in the form of potential wake-words inside a room when the user is not present.
In parallel, it monitors network communications of the devices to detect whether they send data out in response to the audio probe, potentially indicating an audio transmission. 
If a reaction is detected, the probing device notifies the user of the corresponding device's identity.
%about which devices are potentially sending audio out from the user's network. 
%Similarly, by playing back possible wake-words, devices can be tested for unknown wake-words to determine if the device reacts to an utterance it is not supposed to.

%\begin{figure}[ht]
%	\centering
%	\includegraphics[trim=12.5cm 4.5cm 9cm 5cm, clip, width=.50\textwidth]{figures/highlevel.pdf}
%	\caption{Context of our system set-up}
%	\label{fig:highlevel}
%\end{figure}
%\vspace{5cm}

%As discussed in Sect.~\ref{sect:intro}, it is known that audio-controlled devices not only react to designated wake-words, but also to other utterances that erroneously trigger the voice detection functionality of the device. To determine how large this set of misinterpreted utterances in practice is
%Devices may react to multiple voice events, some may not be publicly communicated, to asses to which words a device reacts to 
%we propose a fuzzing method for voice UIs in Sect.~\ref{sect:overview-fuzzing}. Our evaluation on a large set of English words in Sect.~\ref{sect:evaluation} shows, somewhat surprisingly, that voice assistants in practice react to a relatively large set of utterances, many of which do not bear any apparent similarity with the ``official'' wake-words of the system.

%\subsubsection{Voice User Interface}
\ourname offers also a voice-based user interface (UI) for conveniently controlling the probing device and its functions. 
Our current implementation uses a custom Amazon Alexa Skill.
%Controlling the probing device is therefore realized with a Voice User Interface implemented through an Amazon Alexa Custom Skill.
%
%The Skill's backend $L$ runs in the AWS cloud, which has access to a State Server $S$ 
%running on EC2 
%with access to the internet. $S$ is publicly accessible and runs a web server where the probing device $D$ can upload its findings (e.g., list of devices potentially listening to the user during the past hour). Furthermore, $L$ can post to $S$ commands for $D$ to fetch. $D$ is therefore pulling $S$ every 11 minutes for new commands and is simultaneously uploading its results.
%
The Skill has intents and sample utterances for querying the device, e.g., "Alexa, ask \ourname tell me the probing results" and controlling it with, e.g., "Alexa, tell \ourname I'm leaving for one hour".
%For querying, the Skill will start an intent on $L$ which then pulls the stored information from $S$ and
The \ourname then returns a textual response played back using Alexa's voice. 
If the user notifies the system that she is going to be absent, the Skill will start an intent which tells
%$L$ to send a post request to $S$ which then stores it as long as $D$ queries it the next time. $D$ will then, e.g., 
 the device to start probing for the next hour when the user is gone.
With this interface, the user can control when the device is allowed to probe to avoid annoyance while the user is at home. Additionally, the user can query the results of the last probing session to learn which devices responded to the probing and streamed audio to the cloud. We present this Skill-based method as an example of interacting with \ourname. However, optimizing the usability of the user interactions for obtaining the report is not in the core focus of this paper.

\ourname can also be thought of as a complementary component to existing cyber security products like the Firewalla\footnote{https://firewalla.com/} gateway, which gives control and insight over all devices in the network or the Ubiquiti UniFi Wi-Fi access point,\footnote{https://www.ui.com/unifi/unifi-ap/} which can monitor and control connected devices. As such, \ourname could be integrated into already deployed gateway devices.

\end{longversion}

\subsection{Audio Event Detection}
Due to the encryption between devices and back-end cloud systems, it is not possible to detect audio-related events by inspecting packet payloads.
Instead, \ourname identifies audio transmissions by observing sudden outgoing traffic rate increases for specific devices, or significant changes in the device's communication behavior, both of which can indicate the transmission of audio recordings.
%This detection consists of two parts: (1) burst detection, and (2) statistical probing.
We consider two possible complementary approaches to audio event detection: 
%\ourname combines two complementary approaches.
(1)~a simple baseline method based on network traffic burst detection, and
(2)~a statistical approach for detecting apparent changes in the communication characteristics of monitored devices in response to performed audio probes.
\changed{Both audio event detection mechanisms can be used either while actively probing devices,  
%devices to find out if the devices react to audio and to which audio stimuli they react to 
or, without active probing (i.e., when the user is at home) to detect when a device reacts to the noise the user makes (e.g., speaking, playing music) and notifying the user about such activations.}

%\subsubsection{Challenges}
%We also evaluate whether a Machine Learning approach based on known audio traffic of one device is able to recognize audio traffic of other devices. \todocite{keep mentioning of ML here?}

\subsubsection{Burst Detection}
\label{sect:burst-detection}

%\ourname uses burst detection approach as a baseline. \will{what do you mean by "baseline" here? baseline for what?}
Our baseline approach for detecting audio transmissions is based on burst detection in the observed network traffic of devices.
%We first develop an approach for detecting audio transmissions based on burst detection. 
To do this, we need to first identify the characteristics of potential audio transmissions.
We therefore analyzed the invocation process and data communication behavior of popular microphone-enabled IoT devices when they transmit audio. 

%Our burst detection approach for detecting audio transmission is based on monitoring the outgoing communications of a device for an increase in the traffic rate. %~\cite{cheng2018dewicam}.
%if and when devices in a users home react on voice audio under live streaming mode.
%It detects bursts by considering features of network packet headers (e.g., IP address, type, etc.). 
%In our case, the goal is to detect bursts of audio data being sent out by a device in response to a voice-based audio event. 
Our traffic analysis was based on data of popular IoT devices with integrated virtual assistant support: 
(1)~Echo Dot (Amazon Alexa), 
(2)~Google Home (Google Assistant),
(3)~Home Pod (Apple Siri), and
(4)~an audio-activated home security system (Hive Hub 360).
Our analysis showed that these devices do not typically send much traffic during normal standby operation. 
Therefore, it is possible to detect audio transmissions through the increase in traffic rate they cause.
Our approach is generic in the sense that it is applicable to all devices sending audio. We chose these microphone-enabled devices, as they are popular devices produced for a broad range of use cases.
%that used to exchange the information.
%\ourname utilizes burst detection to identify devices potentially reacting to audio events. 
To determine the parameters for burst detection, we monitored the 
%It does so by monitoring for bursts in a device's 
network traffic of devices in response to audio probes emitted into the devices' environment. 
%Detection results obtained during multiple rounds of audio probing are then combined to in order to improve the accuracy and robustness of detection. 
%Based on captured traffic, it identifies a set of suspicious packet flows for further evaluation. When a wireless device sends data out in unauthorized way, BurstDetector will analyze this traffic and inform the user that data was send.

We perform audio event detection by observing sudden increases in the traffic rate emitted by a device that is sustained for a specific amount of time. 
This is because an audio transmission will inevitably cause an increase in the data rate that will typically last at least for the duration of the transmitted audio sample.
This is consistent with how most voice-controlled IoT devices utilizing cloud-based back-ends function (Section~\ref{sect:preliminaries}), where local wake-word detection causes subsequent audio to be streamed to the back-end.
%In a sequence of transmitted data there may be two types of packets: those that represent the sent audio typically transmitted over TLS and those that represent other types of traffic (e.g. ACK, NACK, UDP). Therefore we label all TLS packets as 'potential audio' and all other packets as 'non-audio'. 
% For example, a conversation with Alexa can consist of multiple rounds, e.g. "Alexa, turn down the music", whereupon Alexa responds with "In which room?" and the user replys "In the living room".
%After such audio event the IoT device sends data that consists of packet flow to the server.
%The goal of the burst detection is therefore to detect whether or not the device sent audio data independently from the duration of the conversation.
%In other words, detect the periodic flows of packets that represent exchanged data with server, thus characterize it as audio invocation of the wireless device. To this end, it's not necessary to distinguish all packets that correspond to specific audio event, but enough to specify minimal flow size that corresponds to device invocation that will be enough for transmission detection.

%We denote the minimum number of packets in a flow to be categorized as a possible audio responce with $t_f$ and empirically set $t_f = 80$ based on our testing datasets.
%BurstDetector analyses and characterizes the traffic that consist this two type of packets and informs the user about audio invocation of the IoT device. 

Specifically, \ourname performs burst detection by dividing the network traffic originating from a device into time windows $ W = (w_1, w_2, \ldots) $ of size $ s_w $ and calculating for each time window $ w_i $ the sum of packet payload sizes of the packets falling within the window.
We then calculate the average traffic rate $ B_i $ during the time window $ w_i $ in bytes per second. 
If the traffic rate $ B_i $ is above a threshold $ B_\mathit{audio} $ during at least $ n $ consecutive time windows \[ W_i = (w_i, w_{i+1}, \ldots, w_{i+k-1}) \] where $ k\geq n $, detection is triggered.
%Based on our analysis, we selected $ $. % based on our data collected from several voice-controlled IoT devices (cf. Sect.~\ref{sect:evaluation}). 
Empirically, we found that 
$B_\mathit{audio} = 23kbit/s$ is sufficient to separate audio bursts from background traffic.
%but it is also clearly less than, e.g., the default data rate of ca. $ 256 \si{\kilo}\text{bit} / \si{\second}$ used by Alexa (16-bit PCM at $16\si{\kilo\hertz}$\footnote{https://developer.amazon.com/docs/alexa-voice-service/speechrecognizer.html}) for transmitting user audio
Therefore, it is reasonable to assume our approach will work also for systems using audio codecs with lower bandwidth requirements than Alexa.

\begin{table}[t]
\centering
\caption{Parameters and packet features used by our Burst Detection and Statistical Probing approaches}
\label{fig:features}
\vspace{-1em}
\small
\begin{tabular}{l|l|l}
\thickhline
\textbf{Approach}            & \textbf{Parameters}                 & \textbf{Packet Features}   \\ \hline
Burst Detection     & Window size $s_w$          & Packet size       \\
                    & Traffic rate $B_{audio}$   & MAC/IP            \\
                    & Consecutive detections $n$ &                   \\ \hline
Statistical Probing & Bin count $k$             & Packet size       \\
                    & Packet sequence duration $d$   & Interarrival time \\
                    &  P-value threshold $t$       & MAC/IP            \\
\thickhline
\end{tabular}
\normalsize
\end{table}

As shown in Table~\ref{fig:features}, \ourname uses predefined parameters and packet features. 
%
%From the recorded traffic, 
We extract the packet sizes, the corresponding MAC or IP (depending on the layer), and (implicitly) the direction of the packet (leaving or entering the network). 
To evaluate the optimal consecutive detection threshold $n$ (Section~\ref{sect:bde}), we used a fixed window size $s_w = 1s$, as common voice commands rarely take less than a second. 
For the traffic rate threshold $B_{audio}$, we chose $23kbit/s$. 
This value is a sufficiently low threshold to capture any audio and sufficiently high to discard anything else, as voice recognition services need good voice recording quality (e.g., Alexa's Voice Service uses $ 256 kbit/s$, 16-bit PCM at $16kHz$~\cite{avsaudio}).

%Alg. \ref{alg:BDet} shows the pseudo code of the burst detection process. First a sliding window is initialized, denoted as $sw[M]$.
%to perform the packet analysis.
%Then it analyzes if there are packets possibly containing audio in the sequence, called $packets_t$. If the amount of these packets in the current time frame exceeds the predefined minimal flow size $t_f$, the detector assumes the start of an audio event. During the burst sequence noise packets, defined as $packets_{NT}$, could separate the audio event (e.g. when Alexa is waiting for a response). In this case, we use a counter of permissible packets that could appear between audio packets
%that signify transmission of audio data to the server
%, defined as $NotTLS$, to distinguish audio packets that belong to the same device invocation. If the amount of noise packets is bigger then the permissible amount inside the audio packet sequence, it means that the next sequence of audio packets will belong to the next invocation of the device.

%\begin{algorithm}
%	\caption{BurstDetector}\label{alg:BDet} 
%	\hspace*{\algorithmicindent} \textbf{Input:}{ Packet stream, Bit rate $B$, time frame $tF$} \\
%	\hspace*{\algorithmicindent} \textbf{Output:} {Number of audio transmissions }
%	\begin{algorithmic}[1]
%		\State $time_{0}\gets packet_{0}.time $
%		\While{data streaming}
%		\If{$packet.time <  time_{0} + tF$}
%		\State $PayloadSum = PayloadSum + packet.length$
%		\Else 
%		\If {$PayloadSum > B$}
%		\State audio transmission
%		\Else
%		\State no audio transmission
%		\EndIf
%		\EndIf
%		\EndWhile
%	\end{algorithmic}
%\end{algorithm}

\subsubsection{Statistical Probing}
\label{sect:statistical-probing}
%\subsubsection{Statistical test}

\ourname uses statistical probing to refine audio transmission detection by eliminating cases where traffic bursts result from non-audio transmission.
Most importantly, the approach is generic and does not need \emph{a priori} knowledge of a device's behavior.
It also can determine if a device's communication behavior changes significantly in response to audio probes.
%emitted by the probing device. 

To detect devices that react to audio, we monitor network traffic for significant changes in the device's communication behavior in response to audio probes. 
This is done by determining whether the distribution of the properties of communication packets transmitted by a device after the emission of an audio probe is statistically different from the distribution of packets observed before the probe injection.
Specifically, \ourname uses a $t$-test~\cite{ttest}, which is one of the most commonly used statistical tests.
%, a statistical tool that is used to assess the effect of a IoT device invocation.
Given two data samples, the test computes a $t$-score by determining the data samples' distributions' means and standard deviations, and mapping this to a $p$-value. 
If the $p$-value is below a specified threshold,
% chosen based on an alpha level
 the distributions are considered statistically different and therefore indicate that the device reacted to the audio probe.
%To detect audio invocation, our approach uses an alpha level of 0.05, typical for scientific studies.
\changed{The $p$-value threshold is therefore a system parameter which can be tweaked to produce a trade-off between sensitivity and false alarm rate. However, this threshold is independent of the device type, i.e., a system-wide threshold value is used. The evaluation of this parameter is described in Section~\ref{sect:lpe}.}

First, the probing device monitors idle device traffic while it is not emitting audio probes. It captures a sequence \[ T_s = (pck_1, pck_2, \ldots, pck_n) \] of duration $d$ seconds of data packets $ pck_i $ and calculates a packet size (or inter-arrival time) distribution vector \[ \vec{F_s} = (f_1, f_2, \ldots, f_k) \] by binning the packets $ p_i \in T_s$ into $k$ bins based on the size (or inter-arrival time) of these packets\footnote{To determine the binning automatically, we use \texttt{numpy.histogram} with the bin option \texttt{auto} which uses the maximum of the Sturges and Freedman Diaconis Estimator.} and where $ f_i $ denotes the number of packets assigned to the $i$-th bin.

The probing device then emits multiple audio probes and captures associated data traffic \[ T_{pr} = (pck_1, pck_2, \ldots, pck_n) \] of duration $d$ seconds and extracts a packet size (time) distribution vector $\vec{F_{pr}}$ using the same binning as for $ \vec{F_s} $. The packet size vectors $ \vec{F_s} $ and $ \vec{F_{pr}} $ are then used in the $t$-test to determine a $p$-value ($p$) indicating the likelihood that both frequency samples originate from the same distribution (i.e., from the distribution the device produces while in idle mode). 

If the $ p $-value is below a specified threshold $ t $ (i.e., $p < t$), we assume the traffic samples are not from the same distribution.
That is, the device has reacted in some way and changed its communication behavior. 
To further refine the results, the $p$-value resulting from the packet size distribution is combined with the $p$-value of the inter-arrival time distribution. 
However, as shown in Section~\ref{sect:lpe}, only using the $p$-value of the packet size distribution is sufficient.

%The t-test is invalid when the frequencies given are below 5 so we check for low frequencies and may retest on occurrence.
We collected idle data samples $T_s$ from multiple voice-controlled devices and discovered that they contained frequently occurring smaller data bursts (possibly related to, e.g., heartbeat messages) and infrequently occurring larger data bursts (e.g., state synchronization). 
This observation indicates it is possible to capture a large data burst in one of the two samples ($T_s$ or $T_{pr}$) while missing it in the other. 
Therefore, when encountering a $ p $-value indicating a possible reaction of the IoT device, the probing test must be repeated several times to ensure the observed device behavior change is caused by the audio probe and not by background traffic bursts.
%With that we have three possible outcomes: $B_S$ has much more packets than $B_p$, $B_s$ has much less packets than $B_p$ or both bins have roughly the same amount.

%If it happens that $B_s$ or $B_p$ is much larger than the other we have to scale that bin down to fit the other.
%If it happens that $B_s$ or $B_p$ is much larger than the other we have to retest this wake-word as we don't know in the case of bigger $B_p$ scenario if these bigger bins occurred because of an infrequent big data burst or the device reacting to the possible wake-word. On the other hand if $B_s$ are much larger, e.g. the user could have talked to the device or the device got otherwise activated. The only scenario we don't have to retest is the one where both $B$ are equal in size and the t-test returns a high p-value $p$.

%We also cannot get the frequency and other characteristics of these big data bursts as we want the device to instantly start probing and not capturing data for an extended amount of time before working. Also the device cannot be sure the IoT device is idling, as our device has no microphone for privacy reasons.

%After we calculated $p$ for a possible wake-word for every device, we record and compute new $B'_s$ to compare it with $B'_p$ of the next possible wake-word. We do this to compensate for changes in the behavior of devices, e.g., if a device starts streaming non-audio data into the cloud $B_s$ and $B'_s$ can be very different.
%The p-value is used in the context of null hypothesis testing to quantify how likely it is that the null hypothesis is true.

Table~\ref{fig:features} shows the parameters and packet features used for statistical probing. 
As with burst detection, we extract the packet sizes, the corresponding MAC or IP, and (implicitly) the direction of the packets from the recorded traffic. 
Additionally, we extract packet inter-arrival times. 
As discussed in Section~\ref{sect:lpe}, we set the $p$-value threshold $t$ to be $0.42$-$0.43$ to achieve an optimal precision while fixing the packet sequence duration to $d = 60$ seconds.

\subsubsection{Voice User Interface}
\changed{
\ourname offers also a voice-based user interface (UI) for conveniently controlling the probing device and its functions. 
Our current implementation uses a custom Amazon Alexa Skill.
%Controlling the probing device is therefore realized with a Voice User Interface implemented through an Amazon Alexa Custom Skill.
%
%The Skill's backend $L$ runs in the AWS cloud, which has access to a State Server $S$ 
%running on EC2 
%with access to the internet. $S$ is publicly accessible and runs a web server where the probing device $D$ can upload its findings (e.g., list of devices potentially listening to the user during the past hour). Furthermore, $L$ can post to $S$ commands for $D$ to fetch. $D$ is therefore pulling $S$ every 11 minutes for new commands and is simultaneously uploading its results.
%
%The Skill has intents and sample utterances for querying the device, e.g., "Alexa, ask \ourname tell me the probing results" and controlling it with, e.g., "Alexa, tell \ourname I'm leaving for one hour".
%For querying, the Skill will start an intent on $L$ which then pulls the stored information from $S$ and
%The \ourname then returns a textual response played back using Alexa's voice. 
%If the user notifies the system that she is going to be absent, the Skill will start an intent which tells
%$L$ to send a post request to $S$ which then stores it as long as $D$ queries it the next time. $D$ will then, e.g., 
% the device to start probing for the next hour when the user is gone.
%
With this interface, the user can control when the device is allowed to probe to avoid annoyance while the user is at home. Additionally, the user can query the results of the last probing session to learn which devices responded to the probing and streamed audio to the cloud. We present this Skill-based method as an example of interacting with \ourname. However, optimizing the usability of the user interactions for obtaining the report is not in the core focus of this paper.}

%\todo{start addition}
\subsection{Wake-Word Selection}

Users are not always aware of IoT devices that require a wake-word before transmitting audio recordings to the cloud.
While it is possible enumerate the wake-words for known voice assistants, recent reports of third-party contractors reviewing voice assistant accuracy~\cite{BusinessInsider2019,VRTNews2019,Google2019,TheGuardian2019,Motherboard2019,Bloomberg2019-2} highlight the significance of false voice assistant activation.
Therefore, \ourname identifies other wake-words that will trigger the voice detection. %occur during everyday speech.
Note that this approach is different than using mangled voice samples~\cite{carlini2016hidden, vaidya2015cocaine} or other means to attack the voice recognition process~\cite{schonherr2018adversarial, yuan2018commandersong, carlini2018audio}.
We also do not want to limit \ourname to words sounding similar to the known wake-words in order to confuse the voice assistant~\cite{kumar2018skill, zhang2018understanding}.

Using a full dictionary of the English language is impractical.
It would take roughly 40 days to test a voice assistant with the entire dictionary of 470,000 words~\cite{englishwords} at a speed of one word every seven seconds. 
However, by only testing words with a phoneme count similar to the known wake-word, % trigger the device. 
%Also, a wake-word can be constructed out of two words, so it is unlikely that one word will trigger this compound wake-word. With this in mind it is easy to 
the subset of viable words is manageable.
Our intuition is that a benign device will more likely confuse words with a similar structure.
%
%that can be tested in a manageable amount of time. The idea behind this is that a benign device will obviously mix up words with a similar structure more often than words with an entirely different structure. 
%To assemble the list we 
Therefore, we select all words in a phoneme dictionary~\cite{cmudict} with the same or similar phoneme count than the actual wake-word. 
We also used random words from a simple English word list~\cite{infochimps}. %\footnote{Using Infochimps.com word list: Over 354,000 single words, excluding proper names, acronyms, or compound words and phrases. This list does not exclude archaic words or significant variant spellings.}
These words are spoken using a text-to-speech (TTS) engine.

% Will: the remainder of this is redundant. We discuss the light up ring elsewhere, right?
%Due to the very large number of words to be tested we chose to use a text-to-speech (TTS) engine instead of actual humans reading out words. This allows also to eliminate effects arising from variations in pronunciation and the volume of the audio being played. Entries from this list are fed to a Text-to-Speech (TTS) service to translate them into audio and play back to the voice assistant~\cite{diao2014your, alepis2017monkey} using the loudspeaker of the \ourname device. The determination whether the assistant device reacted to the audio probe is performed by observing the visual indicator (e.g., the prominent ``light ring'' on the Amazon Echo) on the assistant device with the help of a luminosity sensor placed on top of the visual indicator.

%\todo{end addition}
\subsection{System Implementation}

%To realize the \ourname approach (burst detection, statistical audio probing as well as wake-word fuzzing), 
The \ourname probing device injects audio probes into the user's environment and analyzes the resulting device network traffic. 
Our current implementation achieves this functionality using the following hardware set-up.
The probing device consists of a Raspberry Pi 3B~\cite{raspberrypi} connected via Ethernet to the network gateway. 
It is also connected via the headphone jack to a PAM8403~\cite{pam} amplifier board, which is connected to a single generic $3W$ speaker.

To capture network traffic, we use a TP-LINK TL-WN722N~\cite{tlwn} USB Wifi dongle to create a wireless access point using \texttt{hostapd} and \texttt{dnsmasq} as the DHCP server. 
All wireless IoT devices connect to this access point. 
To provide Internet access, we activate packet forwarding between the \texttt{eth} (connected to the network gateway) and \texttt{wlan} interfaces. 
Alternatively the device could sniff Wi-Fi packets without being connected to the network using packet size and MAC address as the features. 
This approach would also work for foreign Wi-Fi networks, as it is not required to have the decrypted traffic, i.e. our device does not need to be connected to that network at all: package size information is sufficient.
%The gateway approach has the benefit of also allowing the device to analyze wired devices, the sniffing approach on the other hand  If our device is implemented in the gateway of the network it can utilize the IP addresses and package size as features. 

Finally, \ourname is written in Python. It uses \texttt{tcpdump} to record packets on the \texttt{wlan} interface. We use Google's text-to-speech (TTS) engine to generate the audio played by the probing device. 

%\subsubsection{Statistical Probing}
%To perform statistical audio probing we configured the Raspberry Pi device to act as an access point 

% It consists of four threads, the \texttt{tcpdump} thread called through \texttt{pexpect} (\texttt{popen} is buffering output and because of that is leading to delays in packet processing), a queue thread responsible for endlessly reading from \texttt{pexpect}'s pipe and writing it into a queue, a thread responsible for playing the probing sounds using \texttt{mpg123} and the main thread containing and endless loop fetching packets from the queue and processing them.

%It is necessary to use a dedicated queue thread because the main thread needs to be able to decide if there is a packet available or not. Querying a pipe leads to a lock until there is a packet on the contrary a queue returns a packet or an Empty exception. The main thread is processing outgoing UDP and TCP packets of internal IPs only. 
%As the bin creator using these estimators can create empty bins we remove them by merging empty bins with non-empty ones.
%The bins are then used for the expected frequencies of the scipy.stats.chisquare test.
%$B_s$ are then used for the expected frequencies of the t-test.

%% file: evaluation.tex
%!TeX root=LeakyPick.tex

This section evaluate \ourname's ability to detect when audio recordings are being streamed to cloud servers. 
%We also evaluate the effectiveness of using audio probes emitted into the environment of the LeakyPick device to discover wake-word-activated devices in the user's home.
%Finally, we also compare the detection accuracy of \ourname's statistical probing, which is device agnostic, to alternative machine learning approaches, which must be trained on specific devices.
%Our evaluation seeks to answer the following research questions.
Specifically, we seek to answer the following research questions.
\begin{enumerate}
    \item[\textbf{RQ1}] What is the detection accuracy of the burst detection and statistical probing approaches used by \ourname?
    
    \item[\textbf{RQ2}] Does audio probing with a wrong wake-word influence the detection accuracy?
    %\ourname correctly not detect audio probes that do not have a wake word.
    
   % How well can emitted audio probes be used to identify audio-activated devices reacting to them by sending audio recordings to a back-end system.\will{Something for online probing. I don't know what we are actually attempting to do}
    
    \item[\textbf{RQ3}] How well does \ourname perform on a real-world dataset?
    
    \item[\textbf{RQ4}] How does \ourname's statistical probing approach compare to machine learning-based approaches?
    
    \item[\textbf{RQ5}] Can \ourname discover unknown wake-words?
\end{enumerate}

%We evaluated \ourname in two ways.
%First, we evaluate the detection accuracy of burst detection and statistical probing.
%Second, we deploy \ourname in a room to evaluate its performance in a realistic environment.
%'s the detection performance, we consider two aspects. For one, in Section~\ref{sect:bde} we evaluate the accuracy of the Burst Detection approach to determine how precisely it can identify outgoing audio transmissions. The statistical audio probing approach of \ourname is evaluated in Section~\ref{sect:lpe} to determine how accurately it is able to identify IoT devices responding to audio events.
%showing that it can distinguish idle traffic from traffic containing data even for samples of just one minute and can detect, e.g., one invocation in a two hour sample. Finally, we evaluate the fuzzing approach in Section~\ref{sect:fe} by fuzzing the wake-word of an Alexa device and showing that it is possible to grasp on which additional words Alexa reacts.

\subsection{Experimental Setup} 

\begin{table}[t]
    \centering
    \caption{Devices used for evaluation}
    \label{tab:devices}
    \vspace{-1em}
    \small
    \begin{tabular}{p{1in}|p{2in}}
    \thickhline
    \textbf{Device Type} & \textbf{Device Name}  \\
    \hline
    Smart Speaker & Echo Dot (Amazon Alexa) \newline
                    Google Home (Google Assistant) \newline
                    Home Pod (Apple Siri) \\
    \hline
    Security System & Hive Hub 360\\
    \hline
    Microphone-Enabled IoT Device & Netatmo Welcome \newline
                                    Netamo Presence \newline
                                    Nest Protect \newline
                                    Hive View \\
    \thickhline
    \end{tabular}
    \normalsize
\end{table}

Our evaluation considers 8 different wireless microphone-enabled IoT devices: 3 smart speakers, 1 security system that detects glass breaking and dogs barking, and 4 microphone-enabled IoT devices, namely the audio event detecting smart IP security cameras Netatmo Welcome, Netatmo Presence and Hive View as well as the smart smoke alarm Nest Protect.
Table~\ref{tab:devices} lists the specific devices.
We now describe our dataset collection and evaluation metrics.
%which were placed in vicinity of the \ourname device performing active audio probing. 
%For evaluation we used voice or audio activated IoT devices from different 
%manufacturers including 
%3 smart speakers Echo Dot (Amazon Alexa), Google Home (Google Assistant) and Home Pod (Apple Siri), one audio security system, i.e., Hive Hub 360 and 4 smart microphone-enabled IoT devices (Netatmo Welcome, Netatmo Presence, Nest Protect and Hive View).
%We tested X different device invocations from 10 different users that have different gender and different accent. 

\subsubsection{Datasets}
\label{sec:dataset}

We used four techniques to collect datasets for our evaluation.
Table~\ref{table:db} overviews these four collection methodologies, as well as to which devices the datasets apply.
The following discussion describes our collection methodology.

\myparagraphi{Idle Datasets}
The idle dataset was collected in an empty office room.
It consists of network traffic collected over six hours during which the device was not actively used and no audio inputs were injected. 
\changed{We also made sure to record at least one occurrence of every traffic pattern the devices produce (e.g., for Echo devices every type of periodic bursts).}

\myparagraphi{Controlled Datasets - Burst Detection}
The controlled datasets for burst detection were collected in an empty office room
while injecting audio probes approximately 100 times for each of the studied devices.
In all cases, the injected probe was the known wake-word for the device in question. The Hive 360 Hub device does not use a wake-word, but is activated by specific noise like dog barking and glass shattering. For this device we therefore used recordings of dog barking sounds to trigger audio transmission.
For each device, three different datasets were collected by varying the wake-word invocation interval between 1, 5, and 10 minutes.

\myparagraphi{Controlled Datasets - Statistical Probing}
The collection of the controlled dataset for statistical probing was performed in a way similar to the burst detection dataset.
However, the experiment collected six datasets for each device. 
Each dataset consisted of six hours of invoking the wake-word at intervals ranging from two minutes to two hours. Thereby resulting in datasets with varying ratios of audio-related and idle background traffic.

\myparagraphi{Online Probing Datasets}
%
%The online probing datasets consist of interleaving time windows of audio probing and no audio probing. The duration of each time window was 1 minute, and
%, so that probing and non-probing alternated at one-minute intervals. 
%during each probing window the wake-word was injected at intervals of 10 seconds.
%
Using live traffic of the 8 different devices listed in Table~\ref{tab:devices}
%, the aforementioned 4 voice assistants and other smart microphone-enabled IoT devices, namely the audio event detecting smart IP security cameras Netatmo Welcome, Netatmo Presence and Hive View as well as the smart smoke alarm Nest Protect,
we randomly selected a set of 50 words out of the 1000 most used words in the English language~\cite{1000words} combined with a list of known wake-words of voice-activated devices as possible wake-words to test. We configured our probing device to alternatingly record silence traffic $T_s$ and probing traffic $T_{pr}$ of one minute duration each for every wake-word in the list. $T_{pr}$ was recorded immediately after the device started playing a word from the list repeating the word every 10 seconds in this minute. 
% of recording.

\myparagraphi{Real-World Datasets}
To create a realistic dataset for evaluation,
we collected data from the three smart speakers over a combined period of 52 days in three different residential environments (houses).
The times for each smart speaker are listed in Table~\ref{table:dbreal}. 
During this time period, humans used the assistants as intended by the manufacturer.

In order to evaluate the accuracy of \ourname, the dataset was labeled by recording the timestamps of when the device was recording audio.
This was accomplished by taking advantage of the visual indicator (e.g., a light ring that glows) that Smart speakers use to alert the user when the voice assistant is activated in response to voice inputs. We therefore automated the labeling process in the real-world environment by creating a small specialized device with a light sensor to measure the visual indicator.
Our device consisted of a Raspberry Pi and a Light Dependent Resistor (LDR) in conjunction with a LM393~\cite{ticomp} analogue-digital comparator. 
The LDR sensor was then attached to the smart speaker's visual indicator and protected from environmental luminosity with an opaque foil. 
This setup allowed the Raspberry Pi to record a precise timestamp each time the device was activated and allowed us to label periods with audio activity in the dataset accordingly.

\begin{table}[t]
	\caption{Datasets for Burst Detection, Statistical Probing, Online Probing and Machine Learning}
	\label{table:db} 
	\vspace{-1em}
	\small
	\begin{tabular}{p{1in}|p{1.0in}|p{0.75in}}
	\thickhline
		\textbf{Dataset}    & \textbf{Frequency}                            & \textbf{Devices}                              \\ \hline
		Idle    	        & -     						                & Echo Dot, \newline Google Home, \newline Home Pod, \newline Hive 360 Hub \\ \hline
		Controlled - \newline Burst Detection         & 1min, 5min, 10min & Echo Dot, \newline Google Home, \newline Home Pod, \newline Hive 360 Hub \\ \hline
		Controlled - \newline Statistical Probing & 2min, 5min, 10min, \newline 30min, 1h, 2h & Echo Dot, \newline Google Home, \newline Home Pod, \newline Hive 360 Hub \\ \hline
		Online Probing      & $10s$ during probing windows    & all, cf. Table~\ref{tab:devices} \\ \hline
		Real-World          & real-world, \newline cf. Table~\ref{table:dbreal}       & Echo Dot, \newline Google Home, \newline Home Pod               \\ 
	\thickhline
	\end{tabular}
	\normalsize
\end{table}

\begin{table}[t]
\caption{Duration of collected data in different residential environments (households) while used by humans}
\vspace{-1em}
\label{table:dbreal} 
\small
\begin{tabular}{l|l|l}
\thickhline
\textbf{Amazon Echo Dot} & \textbf{Google Home} & \textbf{Apple Home Pod} \\ \hline
31d             & 15d         & 15d                \\                                      
\thickhline
\end{tabular}
\normalsize
\end{table}

%Captured period of network traffic over the wireless channels represent packet sequence where each line in the sequence corresponds to one packet. Each line in this sequence contains information about packet: the timestamp of the packet, the address where this packet is coming from, the address where this packet is going to, the protocol name, the length of each packet, additional information about the packet content.

\subsubsection{Evaluation metrics}
We evaluate the performance of our detection approach in terms of true positive rate (TPR) and false positive rate (FPR). The true positive rate is defined as $TPR ={\dfrac{TP}{TP+FN}} $, where TP is true positives and FN false negatives, resulting in the fraction of audio events correctly identified as such. Similarly, false positive rate is defined as $FPR ={\dfrac{FP}{TN+FP}} $, where TN is the true negatives and FP the false positives. It denotes the fraction of non-audio events falsely identified as audio events. Ideally we would like our system to maximize TPR, i.e., the capability to identify devices sending audio to the cloud, while minimizing FPR, i.e., generating as few as possible false detections of audio transmissions.

\subsection{RQ1: Detection Accuracy}
In this section we evaluate the detection accuracy of our two approaches: (1) burst detection and (2) statistical probing.

\subsubsection{Burst Detection}\label{sect:bde}
To evaluate the performance of Burst Detection for detecting audio transmissions, we used the controlled dataset for burst detection (Table~\ref{table:dbreal}) to determine its ability to detect audio transmissions correctly.
%\will{Wait, why aren't we using the controlled dataset for burst detection?}

\begin{figure}[t]
	\centering
	\includegraphics[trim=6.0cm 2.5cm 6.0cm 3cm, clip, width=0.9\columnwidth] {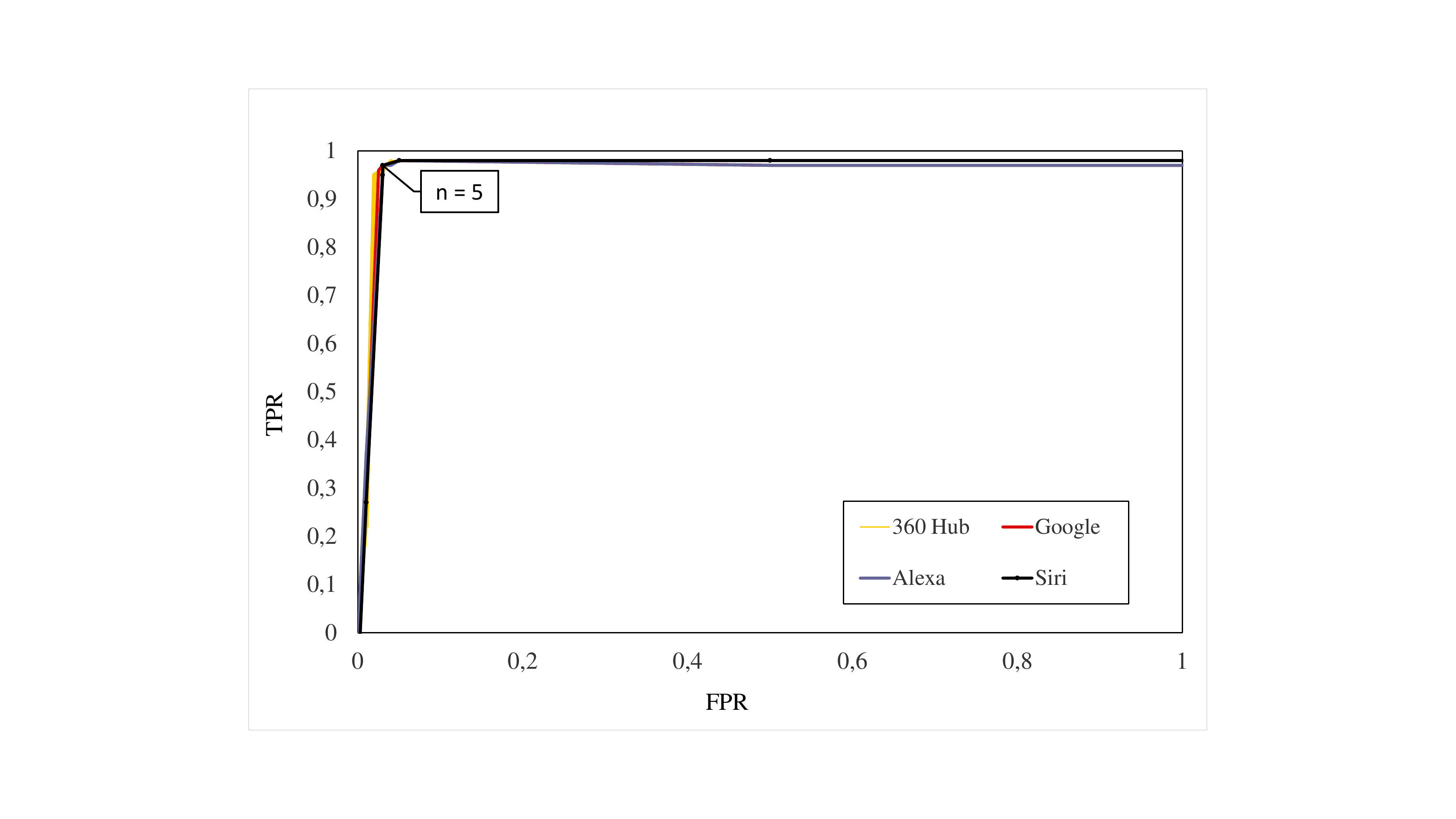}
	\caption{Results of BurstDetector using known wake-words detecting outgoing audio transmissions of Echo Dot, Google Home, Home Pod and Hive 360 Hub on the controlled data set}
	\label{fig:TP-FP}
\end{figure}

Figure~\ref{fig:TP-FP} shows the receiver operating characteristic (ROC) curve for the burst detection approach.
The ROC curve varies the parameter $ n $, which defines the number of consecutive windows with high data throughput required for triggering detection (cf. Section~\ref{sect:burst-detection}), from $ n_{min}=1 $ to $ n_{max}=8 $. As can be seen, with $n$ = 5 consecutive time windows, detection is triggered with a $ TPR $ of 96\% and an $ FPR $ of 4\% (averaged over all devices).
This is explained by the fact that as mentioned in Section~\ref{sect:overview}, the voice-activated devices typically send only a small amount of data unless they are active: medium bursts every few minutes and large bursts only every few hours when idle. %This could lead that during traffic analysis some small amount of false positives caused by this data will be calculated.
%But even considering this 
This allows Burst Detection to identify nearly all audio invocations as they are clearly distinguishable from idle traffic, making this approach practical for devices with such behavioral characteristics.

%\begin{figure}[ht]
%	\centering
%	\includegraphics[trim=5.8cm 1.7cm 6.0cm 2cm, clip, width=0.9\columnwidth] {figures/tpr-fps-4.pdf}
%	\caption{Results of BurstDetector using known wake-words detecting outgoing audio transmissions of Echo Dot, Google Home, Home Pod and Hive 360 Hub on the real world data set}
%	\label{fig:TP-FP}
%\end{figure}
	
\begin{figure}[t]
	\centering
	\includegraphics[trim=8.7cm 4cm 8.7cm 4cm, clip, width=0.9\columnwidth] {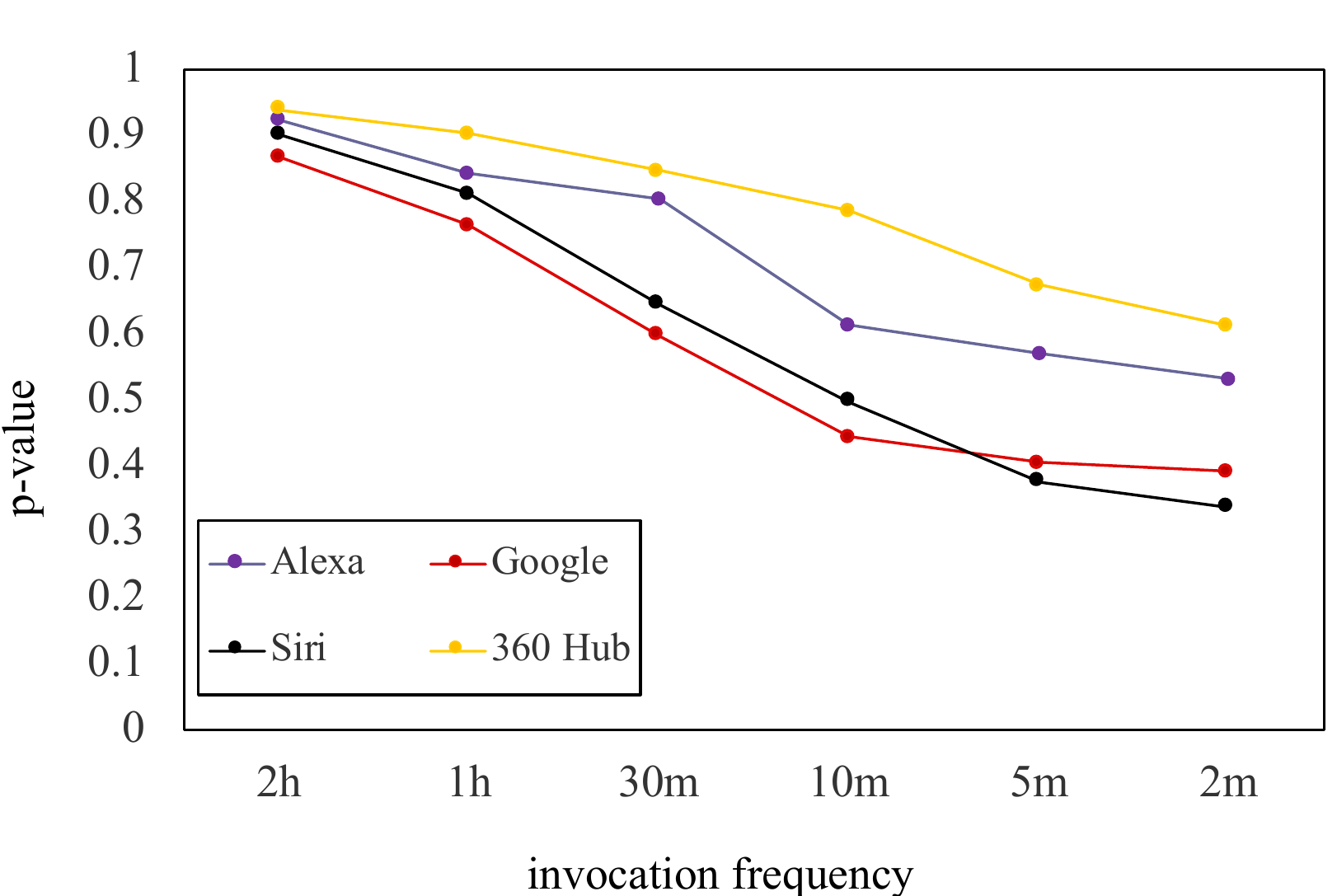}
	\caption{The resulting $p$-value when traffic of devices being invoked in intervals from 2 minutes to 2 hours compared to known silence, showing that the $p$-value decreases with an increasing number of audio bursts in the traffic}
	\label{fig:t-res}
\end{figure}

\subsubsection{Statistical Probing}\label{sect:lpe}
To evaluate the ability of \ourname to detect whether a device reacts to audio events, we first determine whether the statistical properties of data traffic of IoT devices when in idle mode (i.e., not in relation to audio events) is statistically different from the devices' behavior when transmitting audio to the cloud. For this, we calculate the statistical difference of the packet size distributions in the Idle dataset to the packet distributions in the controlled datasets for statistical probing (Table~\ref{table:db}) using the $t$-test as discussed in Section~\ref{sect:statistical-probing}. The results are shown in Figure~\ref{fig:t-res}, showing the resulting $p$-value in dependence of the frequency of invocations of the corresponding device's wake-word. As can be seen, for all tested voice-controlled devices, the $p$-value decreases the more often the wake-word is injected, i.e., the more audio-transmissions the dataset contains. This suggests that the distributions of packet sizes related to audio transmission indeed are different to the distribution of packets in the background traffic and can be thus utilized to identify audio transmissions.

%First, we evaluate the performance of \ourname manually, by recording known idle traffic ($T_{b_s}$) of the studied device (we chose the Alexa Echo Dot) for the baseline $\vec{F_{b_s}}$ vector for the t-test. After that we sample one minute of live silence traffic $T_s$, and extract $\vec{F_s}$. Now we compare the known silence $\vec{F_{b_s}}$ with the assumed silence $\vec{F_s}$. Furthermore we start to sample one minute of probing traffic $T_{pr}$ using the known wake-word of the device (e.g. ``Alexa'') in order to stimulate the device into sending data. $\vec{F_{pr}}$ is also compared to $\vec{F_{b_s}}$ using the t-test.

\begin{figure}[t]
	\centering
	\includegraphics[trim=7.4cm 1.5cm 7.3cm 2.1cm, clip, width=.40\textwidth]{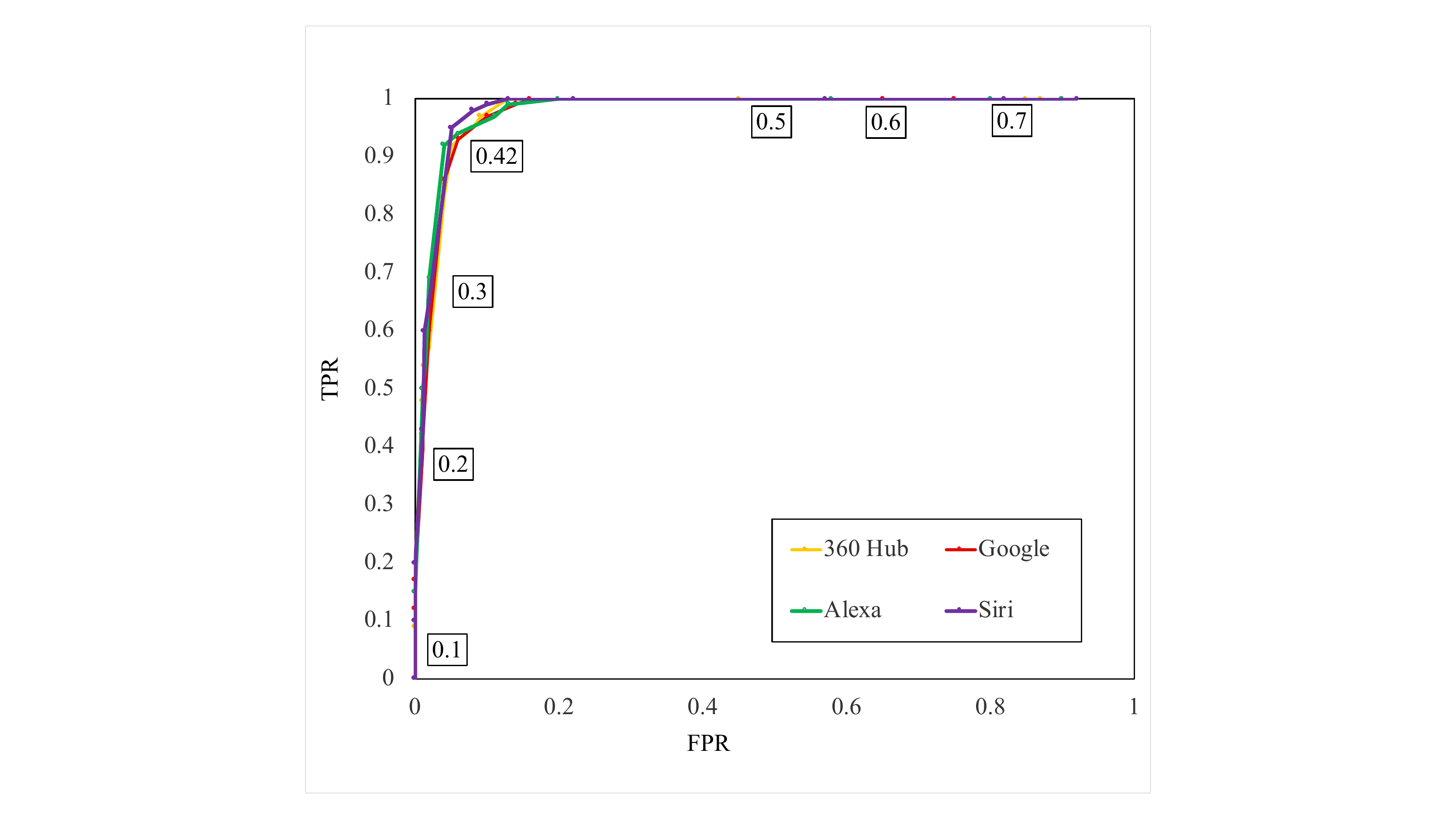}
	\caption{ROC graph of comparing consecutive windows of 30 seconds of traffic of the Controlled - Statistical probing dataset using the $t$-test for different $p$-value thresholds and comparing the output to the actual labels of the traffic}
	\label{fig:finalprobing-roc}
\end{figure}

% Comparing $\vec{F_{b_s}}$ with $\vec{F_s}$ results in p-values with a big variance between around 0.3 and 1.0. This is because the packet count in the one minute idle traffic $T_s$ can be as low as 5 packets, which is not appropriate for a precise output of the t-test. 
Figure~\ref{fig:finalprobing-roc} shows the ROC curve for our approach on the controlled dataset for statistical probing (Table~\ref{table:dbreal}) for different $p$-value thresholds. We use a sliding window approach and compare two consecutive windows of 30 seconds duration using the test, moving the window for 30 seconds to get the new window. We compare the result with the actual label of this traffic region to assess if our approach can reliably find exactly the device sending audio data. As can be seen, for a $p$-value threshold of 0.42 or 0.43 a True Positive Rate of 94\% with a simultaneous False Positive Rate of 6\% averaged over all devices can be achieved for these datasets.
% With this threshold nearly all p-values of invocations are clearly distinguishable from idle, which makes this approach feasible.
%In particular, we evaluate (1) detection accuracy of unexpected audio-based activation events for wireless IoT devices, and (2) the likelihood that particular devices react to specific audio-based keywords. 

\begin{figure}[t]
	\centering
	\includegraphics[trim=8.5cm 5.5cm 5.5cm 1.5cm, clip, width=.50\textwidth]{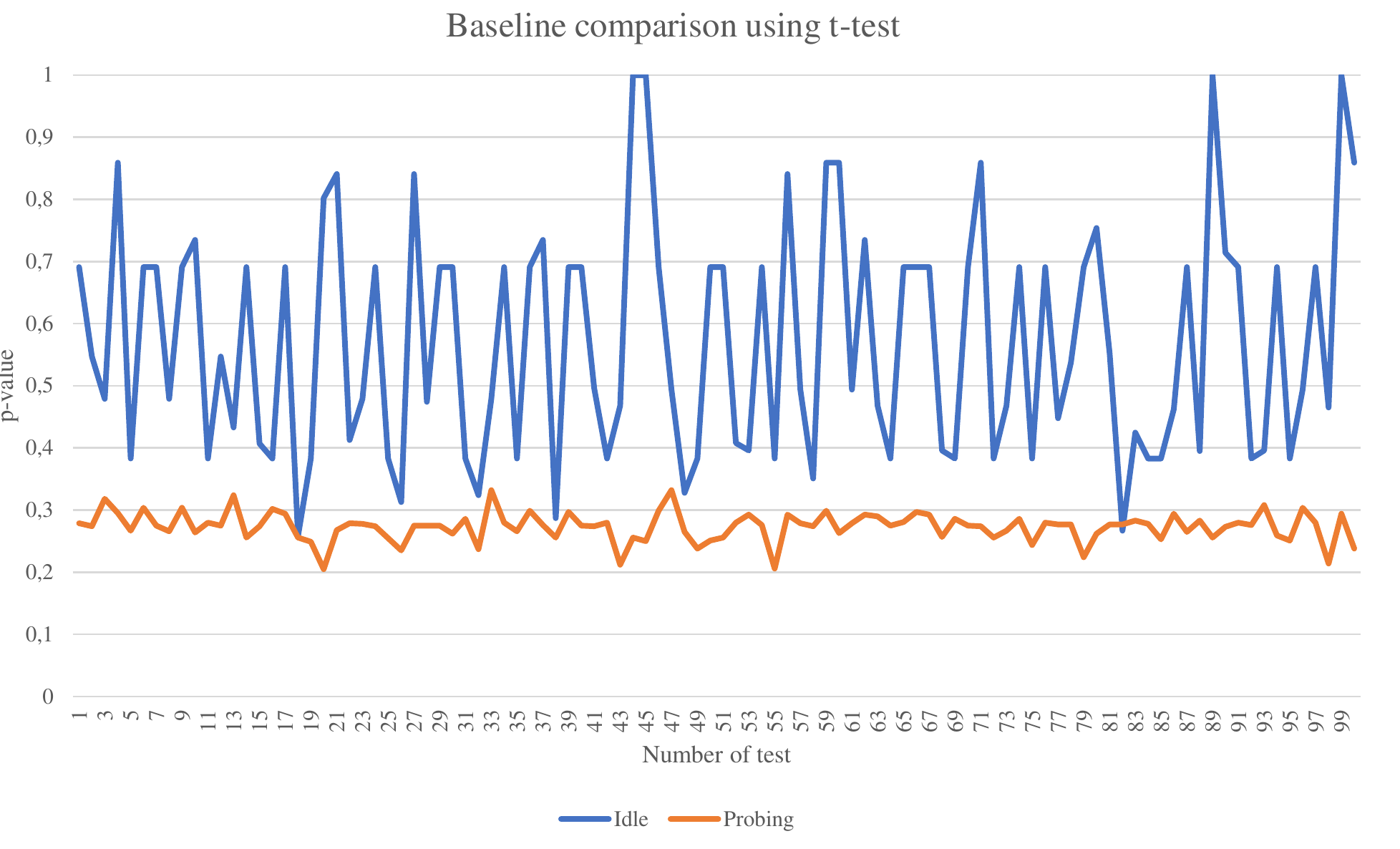}
	\caption{\ourname $t$-test $p$-values for probing Amazon Echo during 100 alternating windows of 1 minute of idle traffic and probing with wake-word ``Alexa'' at 10-second intervals, respectively}
	\label{fig:100ttest}
\end{figure}

\subsection{RQ2: Wake-Word Sensitivity}
\ourname is designed to detect devices reacting to a specific set of wake-words. However, as this set may be different for different device types, a relevant question is to what degree the detection accuracy is dependent on the presence of the correct wake-words in the audio probes.
%\will{We need a more descriptive name than "online probing". It took me a long time to understand what this means}
%\will{What is the research question for this experiment? I'm fairly lost as to what you are trying to show.}
%Now, that we evaluated our two approaches Burst Detection and Statistical Probing, 
To evaluate this aspect, we first tested \ourname on the Online Probing dataset representing a live operative setting in which a number of audio probes containing actual wake-words were injected into the environment of an Amazon Echo device with the target of trying to trigger a wake-word induced audio transmission.
%office environment using audio probing.
%
We used the $t$-test as discussed in Sect.~\ref{sect:statistical-probing} to calculate the $p$-value between consecutive samples of packet sequences $T_s$ and $T_{pr}$ of duration $ d = 60$ seconds each. The result of this for 100 time window pairs is shown in Figure~\ref{fig:100ttest}. As can be seen, the $p$-values for the non-probing (i.e., ``idle'' time windows) range between approximately 0.3 and 1, whereas the $p$-values for time windows containing audio probes remain mostly below 0.3. This shows that given an appropriate $p$-value threshold \ourname is able to distinguish between ``idle'' traffic and audio transmissions.

\begin{figure}[t]
	\centering
	\includegraphics[trim=9.5cm 7cm 9.5cm 5cm, clip, width=.50\textwidth]{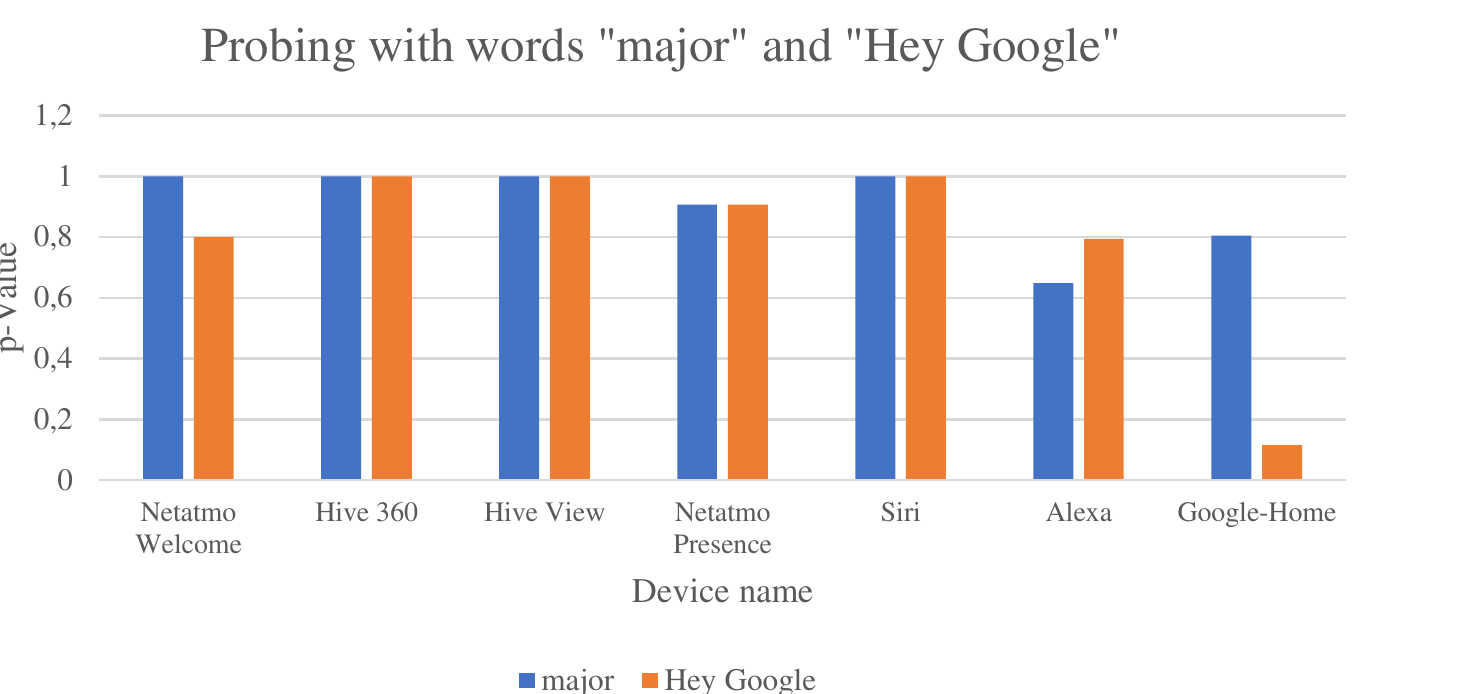}
	\caption{Representative examples of \ourname $p$-values for audio probes. None of the devices react to the non-wake-word probe ``major'' while only the Google Home device shows reaction for its wake-word ``Hey Google''}
	\label{fig:finalprobing}
\end{figure}

To further evaluate how sensitive \ourname is to the use of the right wake-words, we compiled a set of audio probes consisting of the 50 most used English words and a set of nine known wake-words used by the IoT devices used in our evaluation (shown in Table~\ref{tab:devices}). The set of audio probes was injected into the devices' environment and the resulting $p$-values for each device evaluated. \changed{We evaluated all devices at the same time with the same parameters, exactly as it would occur in a smart home scenario where the user has many devices in listening range.}
The resulting $p$-values for two representative examples of used audio probes are shown in Figure~\ref{fig:finalprobing}. The shown audio probes are the randomly-selected word ``major'', which does not correspond to any known wake-word of any of the tested devices and the Google Home wake-word ``Hey Google''. While these examples are provided to demonstrate the discriminative ability of our approach, similar results apply also to other words in the list of tested audio probes.
As one can see, with a $p$-value threshold of, e.g., 0.5 the word "major" would not be considered to activate any of the devices, whereas one can clearly see that the $p$-value for "Hey Google" indicates a reaction by the Google Home device. From the results we can see that only devices responsive to a wake-word react to it which in turn can be detected using the statistical $t$-test employed by \ourname. \changed{This means, that the same $p$-value threshold can be used for any device tested. It shows that only the device actually reacting to the wake word exhibits a low enough $p$-value to be classified as sending audio across all other devices.} Note that Nest Protect is not shown in Figure~\ref{fig:finalprobing}, as it was not activated by any of the examined words and therefore did not transmit any data at all. %This is because Nest Protect runs on batteries and shuts Wi-Fi off to save energy.

%During the course of our evaluation we noticed that the Alexa Echo Dot device did in fact also react to a number of audio probes that were different from the intended wake-word "Alexa", including the words like ``letter'' or ``utter'' when spoken through Google's TTS engine. We are currently investigating whether wrongly recognized wake-words and words in the command part could be used as an attack vector to attack voice assistants and other voice recognition applications. Simultaneously we work on explaining why these wrongly recognized wake-words exist.

%\begin{figure}[ht]
%	\centering
%	\includegraphics[trim=2.5cm 6cm 19cm 6cm, clip, width=.50\textwidth]{figures/finalprobing.pdf}
%	\caption{Results of \ourname using the possible wake-word "major" by comparing one minute silence and one minute probing of all 7 Online devices}
%	\label{fig:finalprobing-major}
%\end{figure}

%\begin{figure}[ht]
%	\centering
%	\includegraphics[trim=19cm 6cm 2.5cm 6cm, clip, width=.50\textwidth]{figures/finalprobing.pdf}
%	\caption{Results of \ourname using the possible wake-word "Hey Google" by comparing one minute silence and one minute probing of all 7 Online devices}
%	\label{fig:finalprobing-google}
%\end{figure}

\subsection{RQ3 and RQ4: Real-World Performance}
\label{sect:traffic-profiling}
%\markus{Start with hypothesis statement}

We evaluated \ourname on our real-world dataset containing 52 days of operation in residential environments (households) (Table~\ref{table:dbreal}).
In addition to using this dataset to measure the accuracy of \ourname~(\textbf{RQ3}), we also compare \ourname's accuracy to that of machine learning algorithms~(\textbf{RQ4}).
Recall from Section~\ref{sect:overview} that a key research challenge is being able to operate for unknown devices.
Since machine learning algorithms require training on known devices, they are not appropriate to achieve our goals, as our approach needs to be able to handle also previously unseen device-types.
That said, we use a trained machine learning algorithm as a baseline, hypothesizing that \ourname can perform at least as well, but without the need for training.

\subsubsection{ML Implementation}
We tested the performance of several commonly-used machine learning (ML) algorithms for detecting audio events in the real-world dataset. We then selected the classifier with the best performance to compare against the statistical detection approach used by \ourname. We consider both simple ML algorithms as well as more advanced ensemble (i.e., Bagging and Boosting) and majority voting-based classifiers. The ML algorithms tested include XGboost~\cite{chen2016xgboost}, Adaboost~\cite{freund1996experiments}, RandomForest~\cite{breiman2001random}, SVM with RBF kernel~\cite{vapnik2013nature}, K-NN~\cite{aha1991instance}, Logistic Regression, Na\"ive Bayes, and Decision Tree classifiers as provided by the the Scikit-learn ML package~\cite{Scikit2020}. For each classifier, the used hyper-parameters were tuned using the provided Grid-search and Cross-validation processes.
%Prior to feeding data to the ML algorithms, we performed 
For constructing features for training we extracted the sequence of packet lengths (SPL) from the traffic flow and utilized the \texttt{tsfresh} tool~\cite{CHRIST201872} that automatically calculates a large number of statistical characteristics from a time-ordered sequence of packets. All experiments were conducted on a laptop that runs Ubuntu Linux 18.04 with an Intel i7-9750H CPU with 32 GB DDR4 Memory.

%This ML approach will work exactly like the Statistical Probing approach described in Sec.~\ref{sect:statistical-probing}. After emitting the audio probed the device will record the traffic sent by the device which is then classified by the ML algorithm.

\subsubsection{Evaluation}
%The Scikit-learn\footnote{https://github.com/scikit-learn/} library was used to implement the machine learning algorithms. 
%(504 data samples) 
%(56 data samples)
For the ML approach,
we used 90\% of the dataset for training and 10\% for testing. In addition, we conducted a 10-fold Cross-Validation (CV) on the training data to better evaluate the performance of the ML classifiers. According to our experiments, based on CV accuracy, the Random Forest Classifier provided the best performance on our dataset, achieving 91.0\%  accuracy (f1-score) on test data while 10-fold CV accuracy was 90.5\%.
%\todocite{Half of this figure is still mock data, as we wait for the real numbers to arrive}
%\begin{figure}[ht]
%	\centering
%	\includegraphics[trim=11.1cm 6cm 11.0cm 6cm, clip, width=0.9\columnwidth] {figures/vsgraph.pdf}
%	\caption{The resulting precision of the \ourname approach and Device Traffic Profiling approaches when using 0.42 as the p-value threshold and training on Alexa labeled traffic}
%	\label{fig:vs}
%\end{figure}

\begin{figure}[t]
	\centering
	\includegraphics[trim=7.5cm 1.5cm 7.5cm 2.2cm, clip, width=.40\textwidth]{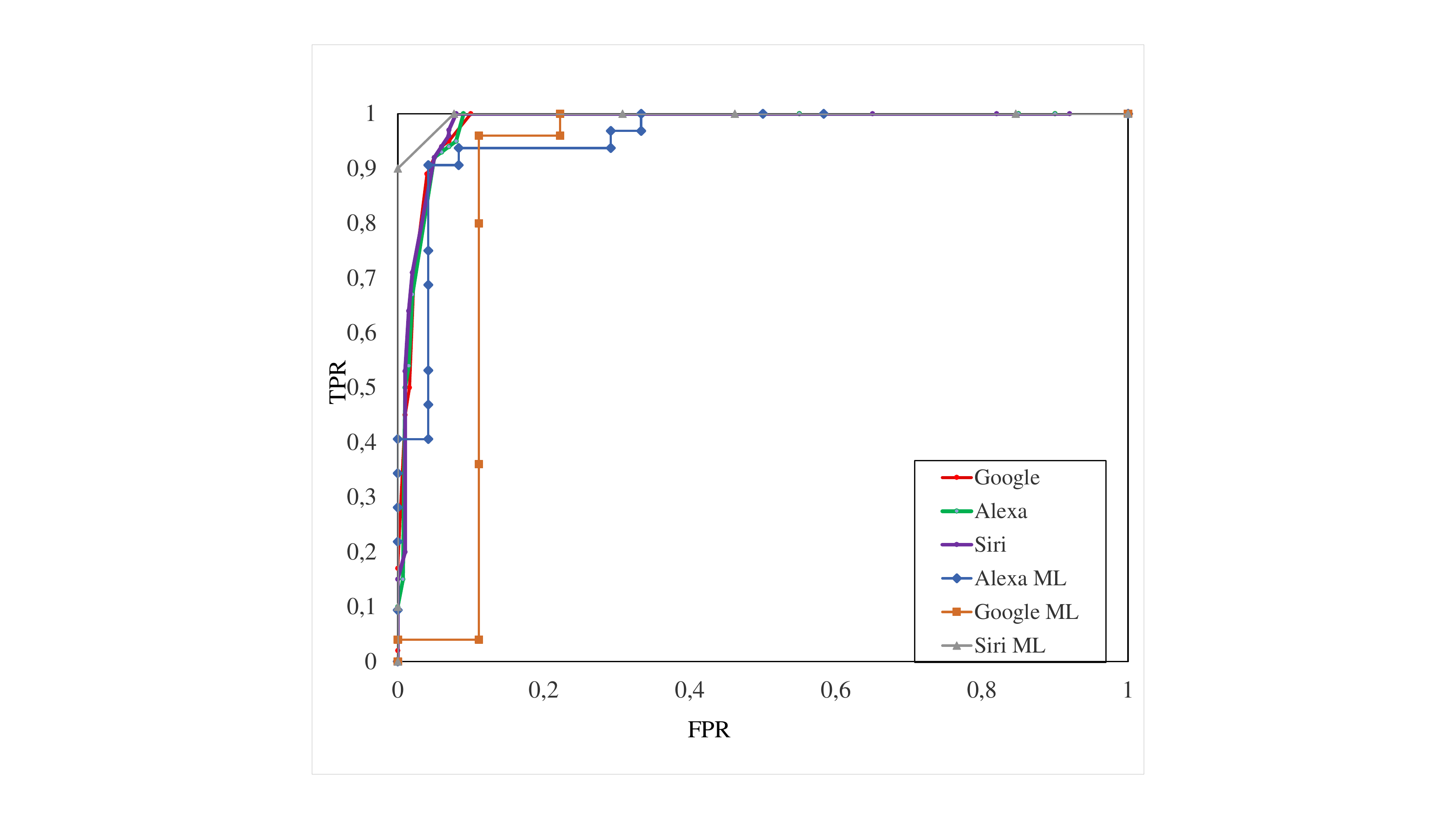}
	\caption{ROC curves of the ML-based and \ourname approaches on the real-world dataset}
	%\caption{ROC graph of comparing consecutive windows of 30 seconds of traffic of the real world dataset using the t-test for different p-value thresholds and comparing the output to the actual labels of the traffic}
	\label{fig:finalprobing-roc-final}
\end{figure}

We also evaluated \ourname as described in Sect.~\ref{sect:lpe} on the same real world dataset in order to compare its performance to the ML-based approach. The results are displayed in Figure~\ref{fig:finalprobing-roc-final}, showing the ROC curves for both approaches on the Google Home, Siri Home Pod and Alexa Echo devices. For $p$-value threshold 0.43 \ourname achieves a TPR of 93\% with a simultaneous FPR of 7\% averaged over all devices, compared to a best-case TPR of 95\% and FPR of 9.7\% for the ML-based classifier for Alexa Echo Dot.
%\todo{start addition}
We also found that models are not transferable between voice assistants.
For example, training on Alexa voice traffic and using the model to identify Siri voice traffic had around 35\% precision.
%While training and validating leads to good results (e.g. on Siri traffic) training a classifier on e.g., Alexa traffic and using it to tell apart Siri voice traffic from idle traffic results in bad performance (around 35\% precision). This shows that the transferability of models between different audio-activated device types is problematic, as the traffic characteristics of
%%, e.g., Alexa devices in comparison to other types of
%voice assistants
%%like Google Home
%are significantly different so that models trained on a particular set of devices likely are unable to detect audio events of other device types reliably.
%\todo{end addition}

As our evaluation results in Figure~\ref{fig:finalprobing-roc-final} show, ML-based models are indeed able to detect audio events based on the traffic the devices send out of the network. However, the evaluation also shows that similar or even better performance can be achieved using a device-agnostic approach as taken by \ourname.

Since applying this kind of a profiling approach requires dedicated traffic models to be trained and included in the system for each device type considered, its practicality in real-world scenarios is questionable. Due to the very large and ever-growing number of different types of voice-activated devices, this approach seems impractical. The approach adopted in \ourname can achieve similar performance without the need to employ pre-trained device-type-specific detection models for audio event detection, providing it much wider applicability in a wider range of environments with diverse audio-activated device types.

%\todo{start addition}
\subsection{RQ5: Identifying Unknown Wake-Words}

To demonstrate \ourname's ability to identify unknown wake-words, we performed a systematic experiment with Amazon's Alexa-enabled Echo Dot.
As voice assistants are conceptually similar, we believe the results can be generalized to other voice-controlled devices.
%We investigated the sensitivity of audio-activation to inputs different than the actual wake-word by performing a systematic experiment with the Alexa-enabled Echo Dot device for finding words other than the configured wake-word, which the device would 'misinterpret' as the wake word and react to them. We chose for our test Amazon's Echo Dot device as it is currently one of the most prominent ones, and the voice recognition of the Amazon system can therefore be assumed to be among the most advanced ones. However, as voice assistants are conceptually similar, we believe that these results can be generalized also to other voice-controlled devices. 
We configured the Echo Dot to use the standard ``Alexa'' wake word (other options include ``Computer'', ``Amazon'', and ``Echo'').
The experiment played different audio inputs, waiting for two seconds for the visual light-ring indicator of the device to light up, indicating the device reacted to the input. 
For each tested audio input, we recorded the number of attempts that triggered a reaction.
Recall from Section~\ref{sect:preliminaries} that Alexa-enabled devices have two states of detection: (1) an offline model on the device, and (2) an online model.
We classify a word to be mistaken as a wake-word when the word triggers at least the offline model, since this transmits recorded audio to the cloud.
%the device is then activated and starts recording and sending audio into the cloud for further processing.

\paragraph{Results} 
The Alexa-enabled Echo Dot reliably reacted to 89 words across multiple rounds of testing.
Table~\ref{tbl:alexa_words} (Appendix) shows the full list of words.
To estimate the phonetic distance between these words and the true wake-word, we used the Metaphone algorithm~\cite{philips1990hanging} to convert the words into a phonetic alphabet based on their pronunciation. 
The resulting words were then compared with the Levenshtein distance to ``Alexa.''
Among the 89 words, 52 have a phonetic distance of 3 or more.
We found that 3 words had a phonetic distance of 0, 9 a distance of 1, 25 a distance of 2, 29 a distance of 3, 14 a distance of 4, 2 a distance of 5 and 6, 4 of 7 and one even a distance of 8. These distances shows that the Echo Dot reliably reacted to words that are phonetically very different than ``Alexa.'' 

\changed{Some of the found wake-words can also be spoken by a human even as part of a sentence and Alexa will be activated. In a smart home scenario users speaking a sentence including such a word can mistakenly activate Alexa and therefore stream the following sentences out of the users home. This shows that those identified words are one cause of misactivations and therefore lead to recorded audio from the users home being sent to the cloud and processed by computers or even other humans.}
Based on these findings, it is unsurprising that Alexa-enabled devices are often triggered unintentionally, leading to private conversations and audio being transmitted outside the user's home.

%However, one can identify groups of similar sounding words among them.
%With these findings we conclude that there are indeed many words that Alexa systematically misinterprets as it's wake-word. 

%It is therefore not surprising that people in a household that utter such words in the vicinity of the Alexa-enabled device can trigger it unintentionally. 
%If triggered the device will start listening and sending the recorded background conversation to the cloud for recognition thus causing unintended audio transmissions to the outside of the user's home. %In addition it may be possible to use these wrong wake-words as an attack vector.%\todo{Should I write about this?}.
%\todo{end addition}

The full results of testing the Alexa wake-word (Alexa) with words of the English language dictionary with 6 and 5 phonemes as well as some random words, is shown in Table~\ref{tbl:alexa_words}. The results shown are the last round of 10 tests for each word. The left column shows the probability of the device being activated while replaying 10 times the word in question.

\input{tables/wake-words.tex}

%% file: tables/wake-words.tex
\begin{table}[t]
\caption{Full results of testing Alexa with English words}
\label{tbl:alexa_words}
\centering
\vspace{-1em}
\begin{tabular}{p{0.30\linewidth}|p{0.65\linewidth}}
\thickhline
\textbf{Probability of \newline activating Alexa} & \textbf{Wake-Word}                                                                                                                                                                                                                                                        \\ \hline
2/10                                                 & alita, baxa, elater, hexer, liker, ochna, taxer                                                                                                                                                                                                                  \\ \hline
3/10                                                 & bertha, electroceramic, excern, oxer, taxir                                                                                                                                                                                                                      \\ \hline
4/10                                                 & electrohydraulic, electropathic, wexler                                                                                                                                                                                                                          \\ \hline
5/10                                                 & blacksher, electic, hoaxer                                                                                                                                                                                                                                       \\ \hline
6/10                                                 & bugsha, elatha, elator, electrodissolution, electrostenolytic, eloper, eluted, fluxer, huerta, hurter, irksome, lecher, lefter, lepre, lesser, letter, licker, lipper, loasa, loker, lotor, lyssa, maloca, maxillar, melosa, meta, metae, muleta, paxar, rickner \\ \hline
7/10                                                 & alexy, crytzer, electroanalytical, hyper, kleckner, lecture, likker, volupte, wexner                                                                                                                                                                             \\ \hline
8/10                                                 & electroreduction, hiper, wechsler                                                                                                                                                                                                                                \\ \hline
9/10                                                 & aleta, alexa, alexia, annection, elatcha, electre, kreitzer                                                                                                                                                                                                      \\ \hline
10/10                                                & alachah, alexipharmic, alexiteric, alissa, alosa, alyssa, barranca, beletter, elector, electra, electroresection, electrotelegraphic, elissa, elixir, gloeckner, lechner, lecter, lictor, lxi, lxx, mixer, olexa, walesa         \\                               
\thickhline
\end{tabular}
\end{table}

%% file: discussion.tex
%!TeX root=LeakyPick.tex

%In this section we propose future work and the discussion for approaches of \ourname.

\myparagraph{Burst Detector}
A malicious audio bug device whose sole purpose is to eavesdrop on a victim may use extensive lossy audio compression to keep the traffic rate below the detection threshold of $23kbit/s$. However, such audio may not be suitable for automated voice recognition as many features of the voice are deleted or exchanged with noise which impairs the scalability of such an attack dramatically. However, our statistical probing approach would still detect a significant difference in the traffic and detect the sent audio.

\myparagraph{Statistical Probing}
As mentioned in Section~\ref{sect:preliminaries}, attacks that issue commands to a victim's voice assistant can be detected by \ourname. To achieve that, increasing the time traffic samples are acquired as well as disabling audio probing is needed. By disabling the audio probing mechanism, every invocation of the device must be done by an external entity (e.g., the user or an attacker). By increasing the sample size, it is also possible to distinguish reliably between an actual invocation and background traffic spikes, even without the knowledge of when audio is played or not as the $p$-values are different for an invocation and background noise (cf. Figure~\ref{fig:t-res}). With this tweak, \ourname would also be able to warn the user of such attacks.
\changed{Currently we are investigating into the influence of varying levels of background noise on the Statistical Probing approach.}

\myparagraph{Countermeasures against Devices sending Audio}\label{sect:counter}
Depending on whether \ourname acts as the gateway of the home network or is sniffing passively the (encrypted) Wi-Fi traffic, there are different approaches to prevent a device from recording and sending audio without the user's permission. If our device is replacing the gateway, traffic identified as containing audio recordings can be simply dropped at the network layer. If our device can only passively sniff encrypted MAC layer traffic, inaudible microphone jamming techniques could be used to prevent the device from recording unsuspecting users private conversations~\cite{roy2017backdoor, song2017inaudible, roy2018inaudible, zhang2017dolphinattack, mitev2019alexa}.

%\todo{start addition}
\myparagraph{Wake-Word Identification}
We found that some of the identified wake-words for Alexa are only effective if spoken by Google's TTS voice, and that we were unable to replicate the effect when spoken by a human. We believe this may result from features that differ between the TTS service audio file and natural voice. 
However, the goal of the experiment was to demonstrate the extent to which words are incorrectly interpreted as wake-words, rather than determining the actual words triggering incorrect wake-word detection.
There may also be wake-words, sounds, or noise our approach could not find. 
% Will: the following sentence is fairly confusing and somewhat contradictory with the previous statements, therefore I'm commenting it out
%However, as we tested the device with a dictionary which includes even rarely used words we assume these missed wake-words to be relatively rarely used therefore posing only a minimal privacy threat. For the frequently used words, our approach can warn the user if such a word triggers the wake-word detector. 
We are currently investigating whether wrongly recognized wake-words 
% Will: not clear how the command part relates to the discussion
%and words in the command part 
could be used to attack voice assistants and other voice recognition applications.
% Will: don't need to say this
%Simultaneously we work on explaining why these wrongly recognized wake-words exist.
%\todo{end addition}

% Reviewer didn't like this addition
\iffalse
\myparagraph{Expected vs. Unexpected Transmission}
%
\morechange{\ourname is designed to detect when a smart home device transmits audio recordings to the cloud in response to observing a sound.
While it does not directly classify if a detected event is expected or unexpected, this determination can usually be determined by the user without significant effort.
For example, if the user does not use any smart speakers, any detected events are unexpected.
If the user has smart speakers, \ourname can provide the MAC and IP addresses of the devices causing the event, which can be compared to the known smart speakers.
Future implementations could automatically filter out known smart speakers.
Finally, for unexpected wake-words for known smart speakers, a known smart speaker list could be annotated with the corresponding voice assistant (possibly automatically by using the MAC address to look up the hardware manufacturer).
Then, only events for wake-words unexpected for that voice assistant need to be brought to the attention of the user.}
\fi

%% file: relatedwork.tex
%!TeX root=LeakyPick.tex

Existing works discussing detection of inadvertent data transmissions out of a user network have focused on IP cameras. To the best of our knowledge, there are no published approaches for detecting outgoing audio traffic for voice assistants and other audio-activated devices, in particular approaches utilizing audio probing. We are also not aware of publications utilizing audio probes to determine if devices react to audio inputs.

The following discussion of related work focuses on existing attacks on voice assistants and traffic analysis approaches for IoT device identification and IP camera detection. We also review approaches to microphone jamming, which can be used by \ourname to prevent microphone-enabled IoT devices to record meaningful audio when the user is not aware of it.

\myparagraph{IP Camera Detection}
IP camera detection approaches usually extract features from packet headers. %the TCP/IP header (e.g. when in-network) or the PHY/MAC header (e.g. by using monitor mode to collect encrypted Wi-Fi traffic).
%The detector makes use of the fact that
Wireless cameras in operation continuously generate camera traffic flows that consist of video and audio streams. The resulting traffic patterns of IP cameras are likely to be different and easily distinguishable from that of other network applications. Furthermore, to save bandwidth, IP cameras utilize variable bit rate (VBR) video compression methods, like H264. Because of the use of VBR, by changing the scene the camera monitors a change in the bitrate of the video can be enforced. Finally, by correlating scene changes and traffic bitrate changes cameras, monitoring can be identified.

Cheng et al.~\cite{cheng2018dewicam} propose using the person being monitored to change the scene by letting them move around. The resulting traffic is then classified using machine learning.
Similarly Liu et al.~\cite{liu2018detecting} focus on altering the light condition of a private space to manipulate the IP camera’s monitored scene. The resulting stream also changes its bitrate and can therefore be distinguished from non-altered streams, e.g., by using the statistical t-test. The above proposals are completely orthogonal to our approach, as they are customized for cameras. In addition, they make assumptions that are not applicable to microphone-enabled IoT devices, e.g., utilizing a variable bit rate encoding (VBR) and continuous data transmission. 

\myparagraph{Traffic Analysis}
Numerous classification techniques have been proposed to learn the behavior of IoT devices, distinguishing and identifying IoT devices based on their traffic profile. Sivanathan et al.~\cite{Sivanathan2017CharacterizingAC} use network traffic analysis to characterize the traffic corresponding to various IoT devices. They use the activity pattern (traffic rate, burstiness, idle duration) and signalling overheads (broadcasts, DNS, NTP) as features to distinguish between IoT and non-IoT traffic. However, the approach requires training. % on traffic recorded for known devices.
Nguyen et al.~\cite{DBLP:journals/corr/abs-1804-07474} propose an autonomous self-learning distributed system for detecting compromised IoT devices. Their system builds on device-type-specific communication profiles without human intervention nor labeled data which are subsequently used to detect anomalous deviations in devices’ communication behavior, potentially caused by malicious adversaries.
However, these proposals focus on detecting anomalous behavior not consistent with benign device actions.
In contrast, our goal is to detect benign actions in response to audio events, which may or may not be falsely detected. 
Also our approach does not require the system to identify IoT devices based on their traffic.
%but will identify devices reacting on audio events. 

\myparagraph{Eavesdropping Avoidance}
Microphone, and more specifically, voice assistant jamming attacks have been proposed by several prior works. Roy et al.~\cite{roy2017backdoor} present an approach for inaudibly injecting audio to jam spy microphones using ultrasonic frequencies and ultrasound modulated noise. As it is inaudible to humans, the jamming is not interfering with human conversations. Zhang et al.~\cite{zhang2017dolphinattack} build upon this work to inaudibly inject commands into voice assistants, demonstrating that voice assistants and possibly other commodity IoT devices are susceptible to the proposed ultrasonic control. Mitev et al.~\cite{mitev2019alexa} further build upon these findings to precisely jam human voice and inject recorded voice into voice assistants.
As discussed in Section~\ref{sect:counter}, inaudible jamming approaches could be used by \ourname to prevent a device from recording meaningful audio when the user is not aware of it. In future work we aim to use these approaches as an additional component of \ourname, further increasing the privacy gains of our approach.

\myparagraph{Voice Assistant Attacks}
Voice assistants using voice recognition are fairly new and many security and privacy aspects are still to be improved.%, leaving room for attacks against them.
The common goal of such attacks is to control the voice assistant of a user without him noticing.%, potentially extracting personal information.
Diao et al.~\cite{diao2014your} present attacks against the Google Voice Search (GVS) app on Android. A malicious app on the smart phone can activate GVS and simultaneously play back a recorded or synthesized command over the built-in speakers which is then picked up by the microphone, to control the victim's voice assistant. 
Alepis et al.~\cite{alepis2017monkey} extend upon this attack. They then proceed to use use multiple devices to overcome implemented countermeasures by showing that infected devices can issue commands to other voice-activated devices such as the Amazon Echo or other smart phones.

%However, as a user standing in vicinity of an attacked device can listen to the recorded or sythesized voice, 
Vaidya et al.~\cite{vaidya2015cocaine} present a method to change a recording of human voice so that it is no longer comprehensible by humans but still correctly recognizable by voice recognition systems. Carlini et al.~\cite{carlini2016hidden} extended this work by presenting voice mangling on a voice recognition system where the underlying mechanics are known, resulting in a more precise attack.
Since a mangled voice may alert nearby users, Schönherr et al.~\cite{schonherr2018adversarial} and Yuan et al.~\cite{yuan2018commandersong} propose methods for hiding commands inside other audio files (e.g., music files) such that they are not recognizable by humans. 
%The commands inside the e.g., music file are only recognizable by voice recognition algorithms and not by humans. 
Similarly, Carlini et al.~\cite{carlini2018audio} create audio files with similar waveforms, which Mozilla’s DeepSpeech interprets as different sentences.

Voice assistant extensions have also been attacked. Kumar et al.~\cite{kumar2018skill} showed that utterances exist such that Alexa's speech-to-text engine systematically misinterprets them. 
Using these findings they proposed \emph{Skill Squatting}, which tricks the user into opening a malicious Skill. 
Simultaneously, Zhang et al.~\cite{zhang2018understanding} proposed using malicious Skills with a similarly pronounced or paraphrased invocation-name to re-route commands meant for that Skill.

These attacks show that an attacker is able to manipulate the interaction flow with a voice assistant by, e.g., issuing commands without the victim noticing, turning voice assistants into a potential privacy and security risk for the user. \ourname can warn the user if their voice assistant is under attack without him noticing it. When combined with eavesdropping avoidance (e.g., jamming), the attacks could be mitigated or even prevented.

%% file: conclusion.tex
%!TeX root=LeakyPick.tex

As smart home IoT devices increasingly adopt microphones, there is a growing need for practical privacy defenses.
In this paper, we presented the \ourname architecture that enables detection of smart home devices that \morechange{unexpectantly} stream recorded audio to the Internet \morechange{in response to observing a sound}.
Conceptually, \ourname periodically ``probes'' other devices in its environment and monitors the subsequent network traffic for statistical patterns that indicate audio transmission.
We built a prototype of \ourname on a Raspberry Pi and demonstrate an accuracy of 94\% in detecting audio transmissions from eight different devices with voice assistant capabilities without any \emph{a priori} training.
It also identified 89 words that could unknowingly trigger an Amazon Echo Dot to transmit audio to the cloud.
As such, \ourname represents a promising approach to mitigate a real threat to smart home privacy.

%We presented and evaluated the \ourname approach. We provide a device which is able to detect whether a device equipped with a microphone is sending audio out of the user's network, whether it reacts to audio inputs or not and to which audio inputs the device reacts. 

%Our solution would increase user awareness of unintended privacy violations of microphone-enabled devices.
%Further, we propose a fuzzing technique which found unpublished and completely different words triggering the wake-word detection functionality of the Alexa voice assistant, demonstrating the extent to which misinterpreted wake words can lead to unwanted transmission of audio from the user's network to the cloud.
%Currently we are working to extend \ourname to work on different malicious settings with devices of different capabilities (e.g., devices deliberately delaying the sending of outgoing traffic).

%Furthermore we suggest limitations of this approach as well as possible modifications as future work.

\begin{acks}
We thank our anonymous reviewers for their valuable and constructive feedback. This work was funded by the Deutsche Forschungsgemeinschaft (DFG) – SFB 1119 – 236615297.
\end{acks}